\documentclass{article}
\pdfoutput=1
\usepackage{arxiv}

\usepackage[utf8]{inputenc} 
\usepackage[T1]{fontenc}    
\usepackage{hyperref}       
\usepackage{url}            
\usepackage{booktabs}       
\usepackage{amsfonts}       
\usepackage{nicefrac}       
\usepackage{microtype}      
\usepackage{lipsum}		
\usepackage{graphicx}
\usepackage{natbib}
\usepackage{doi}

\usepackage{lineno,amsmath,amssymb, siunitx,color, soul}
\modulolinenumbers[1]
\usepackage{booktabs,makecell}
\usepackage{multirow}

\usepackage{xcolor}

\usepackage{wasysym}

\usepackage{subcaption}
\usepackage[english]{babel}
\usepackage[ruled,linesnumbered]{algorithm2e} 
\makeatletter
\newcommand{\removelatexerror}{\let\@latex@error\@gobble}
\makeatother

\usepackage{subeqnarray}

\def\mline{\vrule width3pt height3.0pt depth -2pt}
\def\bdot{\raise.2em\hbox to .15em{.}}
\def\dashed{\mline\hskip4.5pt\mline\hskip4.5pt\mline\thinspace}
\def\solid{\vrule width20pt height2.5pt depth -2pt\thinspace}
\def\dotted{\bdot\ \bdot\ \bdot\ \bdot\thinspace}
\def\chaindot{\mline\ \bdot\ \mline\thinspace}

\title{Data-driven algebraic models of the turbulent Prandtl number for buoyancy-affected flow near a vertical surface}

\author{ \href{https://orcid.org/0000-0001-9194-5998}{\includegraphics[scale=0.06]{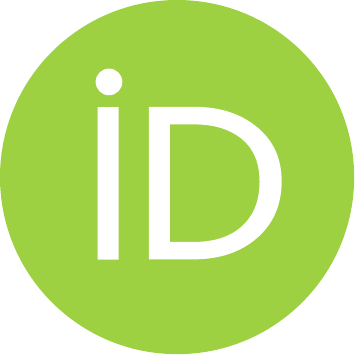}\hspace{1mm}Xiaowei ~Xu}\thanks{Corresponding author.} \\
	Department of Mechanical Engineering\\
	The University of Melbourne\\
	Melbourne 3010, Victoria, Australia \\
	\texttt{xiaoweix2@student.unimelb.edu.au} \\
	\And
	\href{https://orcid.org/0000-0003-1126-7838}{\includegraphics[scale=0.06]{orcid.pdf}\hspace{1mm}Andrew S. H. Ooi} \\
	Department of Mechanical Engineering\\
The University of Melbourne\\
Melbourne 3010, Victoria, Australia \\
\texttt{a.ooi@unimelb.edu.au} \\
	\And
\href{https://orcid.org/0000-0001-5199-3944}{\includegraphics[scale=0.06]{orcid.pdf}\hspace{1mm}Richard D. Sandberg} \\
	Department of Mechanical Engineering\\
The University of Melbourne\\
Melbourne 3010, Victoria, Australia \\
\texttt{richard.sandberg@unimelb.edu.au} \\
}

\renewcommand{\shorttitle}{Data-driven models of turbulent Prandtl number for buoyancy-affected flow}

\hypersetup{
pdftitle={Data-driven algebraic models of the turbulent Prandtl number for buoyancy-affected flow near a vertical surface},
pdfsubject={physics.flu-dyn},
pdfauthor={Xiaowei~Xu, Andrew S. H.~Ooi, Richard D.~Sandberg},
pdfkeywords={Buoyant flow, machine learning, Turbulence modelling, wall-bounded turbulence.},
}

\begin{document}
\maketitle

\begin{abstract}
The behaviour of the turbulent Prandtl number ($Pr_t$) for buoyancy-affected flows near a vertical surface is investigated as an extension study of  {Gibson \& Leslie, \emph{Int. Comm. Heat Mass Transfer},  Vol. 11, pp. 73-84 (1984)}. 
By analysing the location of mean velocity maxima in a differentially heated vertical planar channel, 
we {identify an} {infinity anomaly} for the eddy viscosity $\nu_t$ and the turbulent Prandtl number $Pr_t$, as both terms are divided by the mean velocity gradient according to the standard definition, in vertical buoyant flow.
To predict the quantities of interest, e.g.  the Nusselt number, a machine learning framework via symbolic regression is used with various cost functions, e.g. the mean velocity gradient, with the aid of the latest direct numerical simulation (DNS) dataset for vertical natural and mixed convection. 
The study has yielded two key outcomes: $(i)$  the new machine learnt algebraic models, as the reciprocal of  $Pr_t$,  successfully handle the  infinity issue for both vertical natural and mixed convection; and
$(ii)$ the proposed models with embedded coordinate frame invariance can be conveniently implemented in the Reynolds-averaged scalar equation and are proven to be robust and accurate in the current parameter space, where the Rayleigh number spans from $10^5$ to $10^9 $ for vertical natural convection and the bulk Richardson number $Ri_b $  is in the range of $ 0$ and $ 0.1$ for vertical mixed convection.
\end{abstract}

\keywords{Buoyant flow \and Machine learning  \and Turbulence modelling \and  Wall-bounded turbulence.}

\section{Introduction}
Buoyancy-affected flows near a heated surface have vast engineering applications. Examples include thermal energy systems, e.g. nuclear reactor containment \citep{hanjalic2002one}, building ventilations \citep{batchelor1954heat} and geophysical flows \citep{wells2008geophysical}. 
The turbulent fluid flow for these applications can be numerically simulated using the Reynolds-averaged Navier-Stokes (RANS) equations. 
However, the model-form uncertainties induced by the buoyancy effect have been a long-standing engineering problem.  
One major source of uncertainty is the approximation of  the  turbulence closure terms, which are the Reynolds stress tensor $(-\overline{u_i u_j})$ in the momentum equation and the turbulent heat flux vector $(-\overline{u_i \theta})$ in the temperature equation.
These terms are commonly modelled by the linear eddy viscosity model (LEVM) and standard gradient-diffusion hypothesis (SGDH).  
The bridge between LEVM and SGDH is the turbulence Prandtl number, which is usually defined analogous to the molecular Prandtl number $Pr \equiv {\nu}/{\alpha}$.  For uni-directional flows in a channel:
\begin{linenomath}
	\begin{equation} 
		Pr_t=\frac{\nu_t}{\alpha_t} =  \frac{\overline{uv}}{\overline{v\theta}} \frac{{\mathrm{d} \Theta}/{\mathrm{d} y}}{{\mathrm{d} U}/{\mathrm{d} y}} =  \frac{\overline{uv}}{\overline{v\theta}} \frac{{\Gamma}}{S} , 
		\label{equ:Prt}
	\end{equation}
\end{linenomath}
{where $\nu $ is the molecular viscosity, $\alpha$ is the thermal diffusivity, $\nu_{t} $ is the turbulent eddy viscosity, $\alpha_{t} $ is turbulent thermal diffusivity,  the mean velocity  gradient $S= {\mathrm{d} U}/{\mathrm{d} y} $, and the mean temperature gradient $\Gamma= {\mathrm{d} \Theta}/{\mathrm{d} y}$.
	
	Studies on the turbulent Prandtl number ($Pr_t$) via laboratory experiments, field observations and numerical simulations have a long history.
	Commonly, $Pr_t$ is  treated as a near unity constant (the classical Reynolds analogy, or $\nu_t \simeq \alpha_t$) for air flow as a reasonable approximation.  
	Reviews by \cite{reynolds1975prediction} and \cite{kays1994turbulent} for engineering flows found evidence that  $Pr_t$  deviated from unity in the near-wall region.  Thus, it has been suggested that $Pr_{t}$ should not be a constant but a function of  the distance from the wall or the turbulent Peclet number $Pe_t = (\nu_t /\nu)  Pr$.
	The presence of buoyancy adds complexity to the modelling of the $Pr_{t}$, because of the  increasing dissimilarity between turbulent transport of momentum and heat \citep{li2019turbulent}. The existing literature has mostly paid attention to the flow near a horizontal surface, in which $Pr_{t}$ is modelled according to the  stability conditions caused by heat flux. Popular models include $Pr_t$ as functions of the stability parameter $\zeta = y/L=y/(U^3_\tau/\kappa g\beta \overline{v \theta})$ \citep[][]{monin1954basic}, the gradient Richardson number $R_g ={N^2}/{S^2}$ or flux Richardson number $R_f=\mathcal{G}/\mathcal{P}$  \citep[][]{gibson1978ground, mellor1982development}, where $L$ is the Obukhov length, $N = \sqrt{\left| g\beta \Gamma \right| }$ is  the Brunt–V\"{a}is\"{a}l\"{a} frequency, and $\mathcal{P}$ and $\mathcal{G}$ represent shear production and buoyancy production, respectively, in turbulent kinetic energy (TKE) budgets. Recently, a more unifying framework based on the energy- and flux-budget \citep{zilitinkevich2013hierarchy} or cospectral budget of momentum and heat fluxes \citep{li2015revisiting} has been introduced to analytically formulate the relationship between $Pr^{-1}_t/Pr^{-1}_{t, neu}$ and $R_f$ \citep[see][for a comprehensive review by the atmospheric community]{li2019turbulent}, where the subscript ‘neu’ indicates ‘neutral conditions’.  
	The general observation is that $Pr_{t}$ decreases as the flow become unstable.
	In contrast to the extensive study of flow near a horizontal surface,  the behaviour of $Pr_{t}$ near a vertical surface has rarely drawn attention.
	\cite{gibson1984turbulent}  applied a parametrized model for a vertical setup related to $R_f$  by parametrizing second-moment transport equations that were initially employed for the ground effect near a horizontal surface \citep{gibson1978ground}. 
	It is still not clear whether the aforementioned relations, developed by the atmospheric community, are applicable to the $Pr_{t}$ in vertical setup; 
	nevertheless, the budgets for second-order statistics are different.
	For instance, 
	the buoyancy production $\mathcal{G} = g \beta \overline{u\theta}$ is calculated using the streamwise heat flux for a vertical configuration, whereas $\mathcal{G}$ is based on the wall-normal heat flux  for buoyant flow near a horizontal surface. 
	Furthermore, the existence of velocity maxima ($S\to0$) adds complexity to modelling $Pr_t $ as it tends to infinity.  Therefore, an understanding of the behaviour of $Pr_{t}$ in a vertical setup warrants closer inspection.
	
	There are several vertical configurations \citep{holling2005asymptotic}: flow along a  plate, within an enclosed cavity, along a tube or pipe \citep{jackson1989studies} and that between two infinite differentially heated vertical walls. We choose the latter configuration (see Fig.~\ref{fig:setup}), a fully developed planar channel flow, because of the ideal one-dimensional averaged statistics and the availability of high-fidelity data (either direct numerical simulations (DNSs) or well-resolved large eddy simulations). 
	Seminal works on the vertical setup include the vertical mixed convection (VMC) cases by \cite{kasagi1997direct}  for global Reynolds number $Re_\tau = 150$ with Rayleigh number $Ra = 6.8\times 10^5$ and the vertical natural convection (VNC) cases for $Ra$ at  $\mathcal{O}(10^6)$ \citep{ phillips1996direct,boudjemadi1997budgets,versteegh1999direct}. 
	Recent studies on VMC have focused on analysing the effect of near-wall large-scale structures \citep{fabregat2010identification,wetzel2019buoyancy} using the same parameters.  
	In contrast to the attention on \cite{kasagi1997direct}, 
	the DNS study carried out by \cite{sutherland2015law} at $Re_\tau = 395$ with several $Ra$ cases has received rare attention from modellers.  
	Regarding vertical natural convection,  DNS studies have  more recently extended  the $Ra$  to $\mathcal{O}(10^9)$ \citep{kivs2014natural,ng2015vertical}.  
	This paper will use the newest DNS results \citep{sutherland2015law,ng2015vertical} on the buoyancy-affected vertical channel to develop suitable models. 
	
	Applying machine learning techniques based on high-fidelity data to develop physically informed turbulence models is a burgeoning field \citep{kutz2017deep, duraisamy2019turbulence}.  
	Early studies have applied an optimization method (such as field inversion and the adjoint method) or a Bayesian approach to quantify and reduce the RANS-based uncertainties by modifying turbulent closure terms, model coefficients etc.
	\citet{ling2016analysis} applied a deep neural network method to simple geometrical flows and it showed promising results. Another approach is gene expression programming (GEP) developed by \cite{weatheritt2016novel,weatheritt2017development}. 
	In the comparison of GEP with a deep neural network by \cite{weatheritt2017comparative}, both approaches improved the prediction of the velocity fields for a jet-in-crossflow problem. 
	Recently, similar training frameworks have been implemented for heat flux vector modelling \citep{milani2018machine, milani2020turbulent, sandberg2018applying, weatheritt2020data}. 
	The models are developed by referring to second-order high-fidelity data in an \emph{a priori} sense, called \emph{frozen} training, and the \emph{a posteriori} performance in RANS is sometimes unsatisfactory. 
	Hence,
	a \emph{CFD-driven} training approach \citep{zhao2020rans} was devised to seek better machine learnt candidate models by directly appraising \emph{a posteriori} performance during the training process.
	In this paper, we will utilize GEP with both  \emph{frozen} and \emph{CFD-driven} training to find a proper model for $Pr_t$.

	The primary objective of this paper is a close inspection of the behaviour of the turbulent Prandtl number for buoyancy-affected flow near a vertical surface, 
	which has not seen sufficient attention, 
	since the final study by \cite{gibson1984turbulent}. 
	The paper is organized as follows.  In \textsection \ref{sec:setup}, the location of velocity maxima based on the latest DNS data is shown. We highlight the need for variable $Pr_t$ in the whole domain due to the existence of {an infinity anomaly} for both vertical natural and mixed convection.
	The training framework is then presented in \textsection \ref{sec:gep}, where the detailed procedures of \emph{frozen} and \emph{CFD-driven} training are delineated. 
	Here, we also present the preprocessing method on DNS-based eddy viscosity. In \textsection \ref{sec:res}, the predictive accuracy of GEP-trained models is systematically assessed by investigating the dependency on the training dataset and cost functions. 
	Finally, \textsection \ref{sec:con} concludes this paper. 
	\section{Data source and flow features} 
	\label{sec:setup}
	In this section, we present the setup and the unified governing equations for VNC and VMC. Then, the DNS dataset used in the following modelling process is shown.
	Based on the DNS data, we discuss the behaviour of $Pr_{t}$, which encompasses several distinctive features, {e.g. the existence of a singularity for} the vertical buoyant flow. 
	
	\subsection{Flow setup and governing equation}
	Fig.~\ref{fig:setup} shows a schematic with the three- and two-dimensional view of the setup used in this paper.
	The coordinate system $(x,y,z)$  denotes the streamwise (opposed to gravity direction), wall-normal and spanwise directions.
	When using Reynolds decomposition, the flow instantaneous quantities ($\tilde{u}, \tilde{v}, \tilde{w},\tilde{p}, \tilde{\theta}$) are expressed as the sum of the mean part ($U, V, W,  P, \Theta$) and fluctuations ($u,v, w, p, \theta$). The no-slip and no-penetration boundary conditions are imposed on the velocity and constant isothermal temperatures are set at the walls. 
	Both streamwise ($x$) and spanwise ($z$) directions are periodic for velocity, pressure and temperature. This indicates ${\partial}/{\partial y} \gg {\partial}/{\partial x},  {\partial}/{\partial x}={\partial}/{\partial z}=0, V=W=0 $. Thus, the time- and area- averaged mean profiles $U(y)$ and $\Theta(y)$ only vary along the wall-normal direction. 
	Consequently, the Reynolds-averaged mean equations of motion can be written as:
	\begin{linenomath}
		\begin{align}
			& 0 \simeq -\frac{1}{\rho}\frac{\partial P}{\partial x} + \frac{\mathrm{d}}{\mathrm{d}y}\left(\nu \frac{\mathrm{d}U}{\mathrm{d} y} -\overline{uv}\right) +g_1\beta\left(\Theta-\Theta_0\right), \\
			& 0 \simeq \frac{\mathrm{d}}{\mathrm{d} y}\left(\alpha \frac{\mathrm{d}\Theta}{\mathrm{d} y}-\overline{v\theta}\right).
		\end{align}
	\end{linenomath}
	We treat density $\rho$ as a constant by employing the Oberbeck–Boussinesq approximation for the density variation with temperature in the momentum equations. 
	Furthermore, for simplicity,  we prefer to use $-\overline{u_i u_j}$ instead of $-\rho\overline{u_i u_j}$ and to use $-\overline{u_i \theta}$ instead of $-\rho \overline{u_i \theta}$.
	
	The vertical channel is controlled by two streamwise body forces:
	the gravity force $g_1 =  -g$  and a constant mean pressure gradient $-{1}/{\rho}\:({\partial P}/{\partial x})$.
	The following parameters that dominate the flow, Rayleigh number $Ra$, bulk Reynolds number $Re_b$,  bulk Richardson number $Ri_b$ and Prandtl numbers $Pr \equiv {\nu}/{\kappa}$ 
	are, respectively, defined by,
	\begin{linenomath}
		\begin{equation}
			Ra \equiv \frac{g\beta\Delta \Theta (2h)^3}{\nu \kappa}, 
			Re_b\equiv \frac{2h U_b}{\nu},
			Ri_b \equiv \frac{Ra}{Re^2_b Pr}, \label{eqn:defineRayleigh}
		\end{equation}
	\end{linenomath}
	where the half channel-width is $h$ (full width $H=2h$), $g$ is the gravitational acceleration, 
	bulk mean velocity $U_b=1/(2h)\int_{0}^{2h} U(y)dy$, $\nu$ is the kinematic viscosity and $\kappa$ is the thermal diffusivity. 
	The fluid properties are assumed to be constant.
	The temperature difference $\Delta \Theta = \Theta_h - \Theta_c$ is defined by the scaled temperature $\Theta_h = 0.5$ on the hot plate and $\Theta_c = -0.5$ on the cold plate (see Fig.~\ref{fig:setup}).  It is worth noting that the mean pressure gradient in mixed convection is defined as:
	\begin{linenomath}
		\begin{equation}
			-\frac{1}{\rho} \frac{\partial P}{\partial x}=\frac{U^2_{\tau,h}+U^2_{\tau,c}}{2  h}, 
		\end{equation}
	\end{linenomath}
	where the friction velocities are  $U_{\tau,h} = \sqrt{\nu\: \mathrm{d} U/\mathrm{d}y|_{w, h}}$ at the hot wall and $U_{\tau,c} = \sqrt{\nu\: \mathrm{d} U/\mathrm{d}y|_{w, c}}$ at the cold wall. In addition, the global mean friction velocity is defined by the arithmetic mean of the one at each wall, that is,  $U_{\tau}=\left({U_{\tau,h}+U_{\tau,c}}\right)/{2}$.
	For forced convection, $U_\tau = U_{\tau,h} = U_{\tau,c}$, and $(-{1}/{\rho})\:{\partial P}/{\partial x}=U^2_\tau/h$.
	Lastly, 
	the Nusselt number, the dimensionless heat transfer rate, is quantified as  
	$Nu \equiv {f_w (2h)}/{(\Delta \Theta \kappa)}$ 
	where $f_w \equiv \kappa |{{\mathrm{d} \Theta}/{\mathrm{d} y}}|_w$, $|{{\mathrm{d} \Theta}/{\mathrm{d} y}}|_w$ is the mean temperature gradient at the hot and cold walls.
	
	There are two limit states for the current setup, namely, pure buoyancy-driven flow (referred to as natural or free convection) and pure shear-driven flow  (referred to as forced convection), which can be quantified by the bulk Richardson number $Ri_b$. Increasing $Ri_b$ means adding a buoyancy effect;  if $Ri_b = 0$, it means $g=0$, and the flow is the canonical channel flow (forced convection), while as $Ri_b \to \infty$ (namely, $U_b = 0$), the flow is purely buoyancy-driven (natural convection). Fig.~\ref{fig:setup} ($b$) is a two-dimensional view of the vertical channel, from left to right: natural, mixed and forced convection. In each panel, the mean velocity $U(y)$ and mean temperature $\Theta(y)$ are plotted to show the velocity maxima for the three scenarios.
	\begin{figure*}
		\begin{center}
			\includegraphics[width=0.90\textwidth]{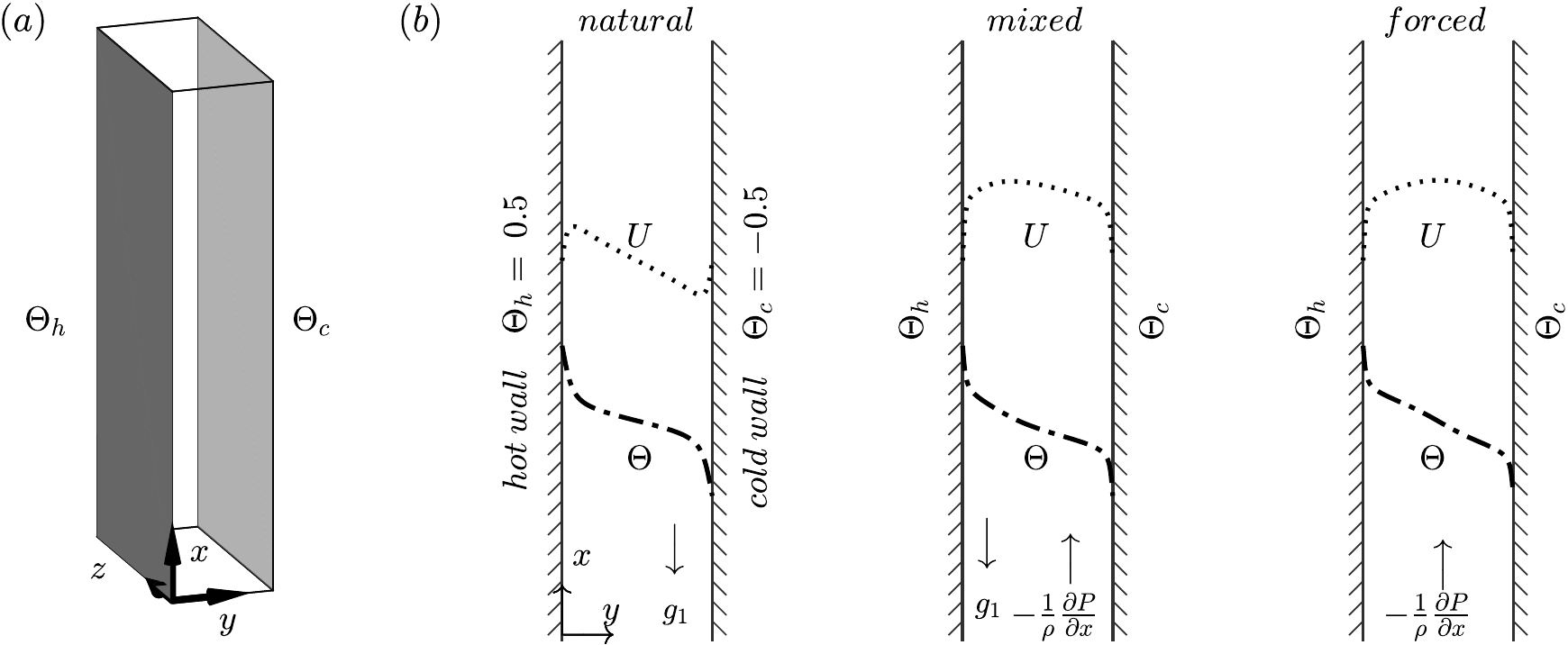}
			\caption{A schematic diagram of the setup of an infinite vertical channel, $(a)$ three-dimensional view of domain used for DNS; $(b)$  two-dimensional view with the shape of mean velocity $U(y)$ and temperature $\Theta(y)$ profiles, from left to right: natural convection (solely driven by gravitation force $g_1$), mixed convection (driven by a combination of $g_1$ and a constant mean pressure gradient $-\frac{1}{\rho}\frac{\partial P}{\partial x}$) and forced convection (solely driven by $-\frac{1}{\rho}\frac{\partial P}{\partial x}$). }
			\label{fig:setup}
		\end{center}
	\end{figure*}
	
	This paper employs different nondimensionalizations for natural, mixed or forced convection due to the distinctive features of shear-dominant and  buoyancy-dominant flows.  For natural convection, we choose the traditional full width $H$ as the length scale  and $H/U_f$ (where the velocity scale is the free fall velocity $U_f= \sqrt{g\beta \Delta \Theta H}$ \citep{ng2015vertical}) as time scale. For forced and mixed convection, the length scale is  the half channel-width $h$, and the time scale is $h/U_\tau$. Note that the mechanical turbulent dissipation rate $\varepsilon$  is normalized by $U^3_f/H$ for VNC and $U^4_\tau/\nu$ for VMC.
	\subsection{DNS dataset}
	In this study, 14 cases of the DNS dataset  are used (see Table~\ref{tab:database}), which were carried out by  \cite{ng2015vertical} (Set A, Case $1\sim7$) for VNC and \cite{sutherland2015law} (Set B, Case $8\sim14$) for VMC with $Pr = 0.709$ (for air flow). The cases cover the range of Reynolds and Rayleigh numbers 
	$0 \le  Re_b \le  1.471 \times 10^4$, 
	$10^5 \le Ra \le 10^9$. 
	We adopt the label \emph{Rax\_Rey} \citep{pirozzoli2017mixed} at $Ra = 10^x$, $Re_b = 10^y$.
	For instance, the flow case 11, {$Ra6.5\_Re4.2$}, denotes $ Ra=  3.6 \times 10^6 = 10^{6.5}$ and $ Re_b =  1.471 \times 10^4 = 10^{4.2}$.
	Besides,  $Ra = 0$ corresponding to pure Poiseuille flow (forced convection), and $Re_b = 0$ corresponding to VNC. 
	Table~{\ref{tab:database}} also provides a shorthand label, for instance, ${Ra80}$ for the natural convection case at $Ra = 1.0\times 10^8$ and ${Ri50}$ for the mixed convection case at $Ri_b=0.050$. 
	\begin{table*}[!ht]
		\centering
		\begin{tabular}{p{0.35cm} c l l c c c c c l}
			\toprule
			& Case	& Label	& Flow case & $Ri_b$ & $Ra$& $Re_b$&  $Nu$ & Flow type	& Purposes	\\[+0.2em]
			\hline
						\multirow{7}{*}{${A}\left\{ \rule{0cm}{1.4cm}\right.$}
			&1	& Ra50	&Ra5\_Re0        &  $\infty$       & $1.0\times10^5$  &  0  &   57.53  & natural & testing \\  
			&2	& Ra57	&Ra5.7\_Re0      &  $\infty$       & $5.4\times10^5$&  0  &   29.18  & natural &  training \& testing   \\ 
			&3	& Ra63	&Ra6.3\_Re0      &  $\infty$       & $2.0\times10^6$  &  0  &   16.64  &  natural &  testing   \\  
			&4	& Ra66  &Ra6.6\_Re0      &  $\infty$       & $5.0\times10^6$  &  0  &   10.82   &  natural &  testing  \\  
			&5	& Ra73	&Ra7.3\_Re0        &  $\infty$       & $2.0\times10^7$  &  0  &   8.15    &  natural &  training \&  testing  \\  
			&6	& Ra80	&Ra8\_Re0        &  $\infty$       & $1.0\times10^8$  &  0  &   5.37    & natural &   testing  \\ 
			&7	& Ra90	&Ra9\_Re0        &  $\infty$       & $1.0\times10^9$  &  0  &   3.04    &  natural &   training \& testing \\[+0.4em]
						\multirow{7}{*}{${B}\left\{ \rule{0cm}{1.4cm}\right.$}
			&8	& Ri00	&Ra0\_Re4.1     &  $0  $     & $0$   &  13846  &   12.75  & forced & training \& testing \\  
			&9	& Ri13	&Ra6.3\_Re4.2   &  $0.013$       & $1.9\times10^6$&  14239  &   12.79  & mixed &  testing   \\ 
			&10	& Ri18	&Ra6.3\_Re4.1   &  $0.018$       & $2.2\times10^6$  &  12963  & 12.32  &  mixed &  testing   \\  
			&11	& Ri23 &Ra6.5\_Re4.2   &  $0.023$       & $3.6\times10^6$  &  14710  & 13.57   &  mixed &  testing  \\  
			&12	& Ri35	&Ra6.6\_Re4.1   &  $0.035$       & $4.0\times10^6$  &  12696  & 12.86   &  mixed &    testing  \\  
			&13	& Ri50	&Ra6.9\_Re4.2   &  $0.050$       & $8.3\times10^6$  & 15232  & 14.88   & mixed &   training \& testing  \\ 
			&14	& Ri94	&Ra6.9\_Re4.2   &  $0.094$       & $9.3\times10^6$  &  11825  & 13.54    &  mixed &   training \& testing \\ 
			\bottomrule
		\end{tabular}
		\caption{DNS dataset of flow cases, $Re_b = 2 h U_b / \nu$ is the bulk Reynolds number, $Ri_b=2 \beta g \Delta \Theta h / U_b^2$ is the bulk Richardson number, $Ray=\beta g \Delta \Theta (2 h)^3 / (\alpha \nu)$ is the Rayleigh number, $Nu= (2h/\Delta\Theta)|{{\mathrm{d} \Theta}/{\mathrm{d} y}}|_w$ is the Nusselt number. Set A, Case $1\sim7$, \cite{ng2015vertical}; Set B, Case $9\sim14$, \cite{sutherland2015law}; more details are showed in Appendix $A$. In the \emph{Purposes} column, the training and testing datasets are showed, in which the \emph{training} represents the cases used to train a machine learnt model, and \emph{testing} represents the cases used for cross-validation. }
		\label{tab:database}
	\end{table*}
	
	\subsection{The behaviour of the turbulent Prandtl number in global coordinates}
	\label{sec:inf}
	For the one-dimensional mean flow field, the governing equations only include the two components of the Reynold stress tensor and heat flux vector, which are the Reynolds shear stress ($\overline{uv}$) and the wall-normal heat flux ($\overline{v\theta}$) for a planar channel flow. We adopted the well-established linear gradient LEVM and SGDH models as the starting point, which are, respectively, 
	\begin{linenomath}
		\begin{align}
			& -\overline{uv} = \nu_t \frac{\mathrm{d} U}{\mathrm{d}y}, \\
			&  -\overline{v \theta} = \alpha_{t}  \frac{\mathrm{d}\Theta}{\mathrm{d}y} =\frac{\nu_t}{Pr_{t}} \frac{\mathrm{d}\Theta}{\mathrm{d}y}.
		\end{align}
	\end{linenomath}
	
	First and foremost, it is essential to discuss the signature of the mean velocity gradient ${{\mathrm{d} U}/{\mathrm{d} y}}$ and  mean temperature gradient ${{\mathrm{d} \Theta}/{\mathrm{d} y}}$, $\overline{uv}  ,\overline{v\theta} $ in global coordinates. \cite{gibson1984turbulent} listed the major features along the vertical surface; they are mostly true only if the flow is in or near the buoyancy-driven/buoyancy-dominated regime. From DNS studies  \citep[see][Fig.~\ref{fig:vnc_loc} $(c)$]{versteegh1999direct}  on the vertical natural convection, the adapted version is, 
	\begin{figure*}[!ht]
		\begin{center}
			\includegraphics[width=1.0\textwidth]{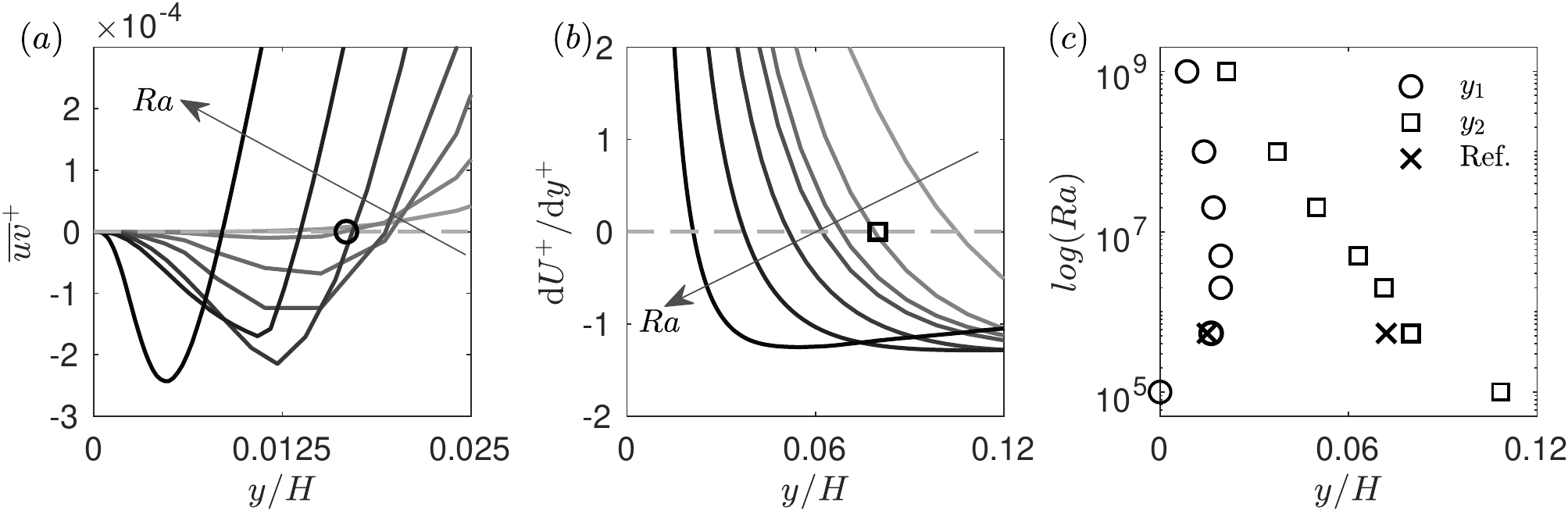}
			\caption{Division of region in VNC: $(a)$, $y_1$ is the zero point of $\overline{uv}$, and $(b)$, $y_2$ is the zero point of ${{\mathrm{d} U}/{\mathrm{d} y}}$ in the region between the hot wall ($y/H=0$) and the centreline ($y/H = 0.5$). The greyscale from light to dark that varies with the increase in $Ra$ from $10^5$ to $10^9$ is based on \cite{ng2015vertical}; see Table~\ref{tab:database} for details on DNS cases.  The $y_1$ and $y_2$ values for $Ra = 5.4 \times 10^5$ are explicitly shown in $(a)$ and $(b)$. $(c)$, the variation in  $y_1$ and $y_2$ with the change in Ra, where the \emph{Ref. $\times$} is a verification case in \cite{versteegh1999direct} at $Ra = 5.4 \times 10^5$.}
			\label{fig:vnc_loc}
		\end{center}
	\end{figure*}
	\begin{itemize}
		\item[($i$)]The temperature gradient (${{\mathrm{d} \Theta}/{\mathrm{d} y}}$) is negative everywhere, and the wall-normal heat flux ($\overline{v\theta}$) is positive everywhere; 
		\item[($ii$)] The Reynolds shear stress  ($\overline{uv}$)  is negative for small $y_w$ and positive for large $y_w$, where $y_w$ is the nearest distance from the wall (or wall distance).
		\item[($iii$)] The turbulent Prandtl number $Pr_{t}$ has singularities and a negative region in the vicinity of the wall.  
	\end{itemize}
	
	For a clear discussion regarding the division of the regions, we define $y_1$ as the zero point of $\overline{uv}$ and  $y_2$ as the zero point of ${{\mathrm{d} U}/{\mathrm{d} y}}$ in the region of the hot wall ($y=0$) to the centerline ($y = h$). 
	Fig.~\ref{fig:vnc_loc} $(a)$ and $(b)$ shows the exact position of $y_1$ and $y_2$, which results in the sign of key quantities (see Table~\ref{tab:sign}). 
	For the whole domain in global coordinates, a laminar sublayer exists within $ 0<y_w<y_1$ and a turbulent layer within $y_2<y_w<h$. The $y_1<y_w< y_2$ region is actually the bridge between these two distinctive regions, which can be called the adjustment region \citep{wells2008geophysical}. 
	Fig.~\ref{fig:vnc_loc} $(c)$ shows that the adjustment region shrinks with the increase in $Ra$ and might diminish when $Ra \to \infty$  \citep{holling2005asymptotic}.
	Meanwhile, \cite{ng2017changes} suggest that the thermal and viscous boundary layers undergo a transition from a classical laminar-like state to the ultimate shear-dominated state from moderate to high $Ra$. 
	Therefore, given the diminishing of the adjustment region and the transition of the laminar sublayer, the whole domain could be turbulent in the ultimate state. 
	From the point of view of the modeller, the adjustment region is the area where the  LEVM \citep{versteegh1999direct} is not valid near both walls. 
	In other words, the infinity issue occurs in both $Pr_{t}$ (see Eq.~{\ref{equ:Prt}} and Fig.~{\ref{fig:vnc_prt}} $(b)$) and $\nu^{dns}_t = -\overline{uv}/(\mathrm{d}{U}/\mathrm{d} y)$ (see Fig.~{\ref{fig:vnc_prt}}$(c)$) in global coordinates.
	Collectively, the infinity issue at $Ra$ from $10^5$ to $10^9$ is a challenging question for symbolic regression and model generality in this study.  
	\begin{table*}[!ht]
		\centering 
		\begin{tabular}{c c c c c c c c} 
			\toprule
			region & ${{\mathrm{d} U}/{\mathrm{d} y}}$   &  ${{\mathrm{d} \Theta}/{\mathrm{d} y}}$ &$\overline{uv}$  &$\overline{v\theta}$  & $\nu_t$ &$Pr_t$\\ [1ex]
			\hline
			$ 0<y_w<y_1$&  $ + $  & $- $ & $-$ & $+$  &  $+$ & $+$\\  
			$y_1<y_w< y_2$ &  $ + $  & $ - $  & $+$ & $+$ & $-$ &$-$ \\ 
			$y_2<y_w<h$  &  $ - $  & $ - $ & $+$  & $+$  &  $+$ &$+$\\ 
			\bottomrule 	 
		\end{tabular}
		\caption{The sign of quantities along wall-normal direction until the centreline in global coordinates for VNC.}
		\label{tab:sign}
	\end{table*}
	\begin{figure*}[!ht]
		\begin{center}
			\includegraphics[width=1.\textwidth]{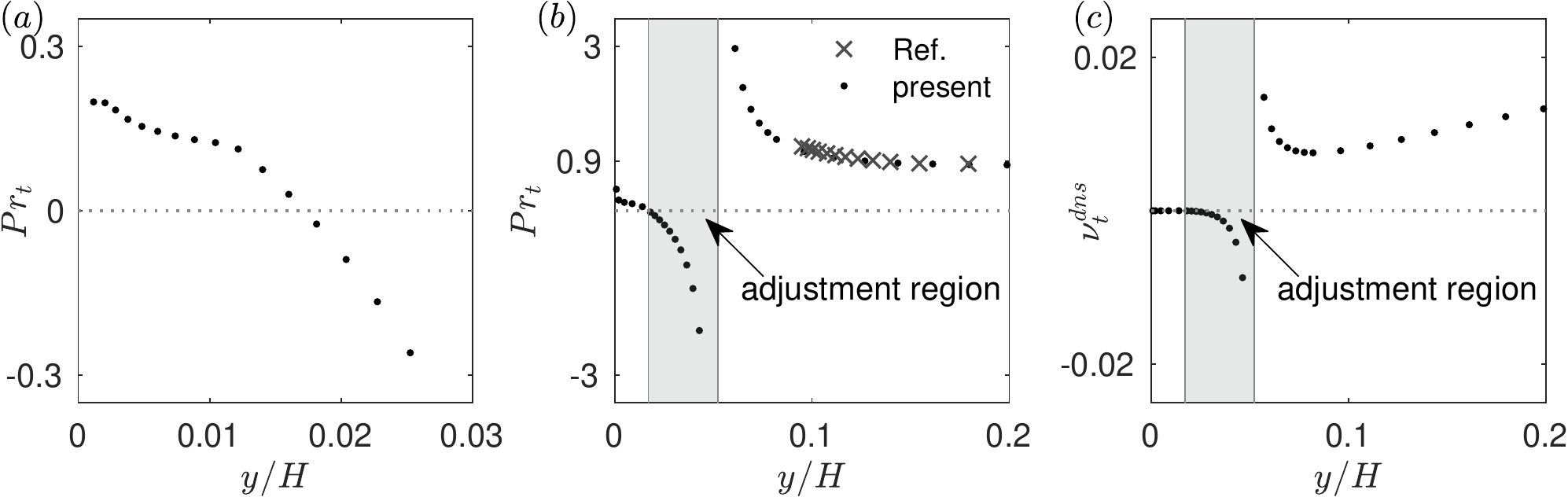}
			\caption{The distribution of  turbulent Prandtl number ($Pr_{t}$)  and eddy viscosity $\nu^{dns}_t$ at $Ra = 5.0 \times 10^6$ for VNC, where $(a)$, $Pr_{t}$ in the near-wall positive region; $(b)$, $Pr_{t}$ in global coordinates, where the present data are validated against \cite{dol1999dns} ($\times$); $(c)$, $\nu^{dns}_t$ in global coordinates. The grey patch depicts the {adjustment} region between $y_1$ and $y_2$.}
			\label{fig:vnc_prt}
		\end{center}
	\end{figure*}
	\begin{figure*}[!ht]
		\begin{center}
			\includegraphics[width=1.\textwidth]{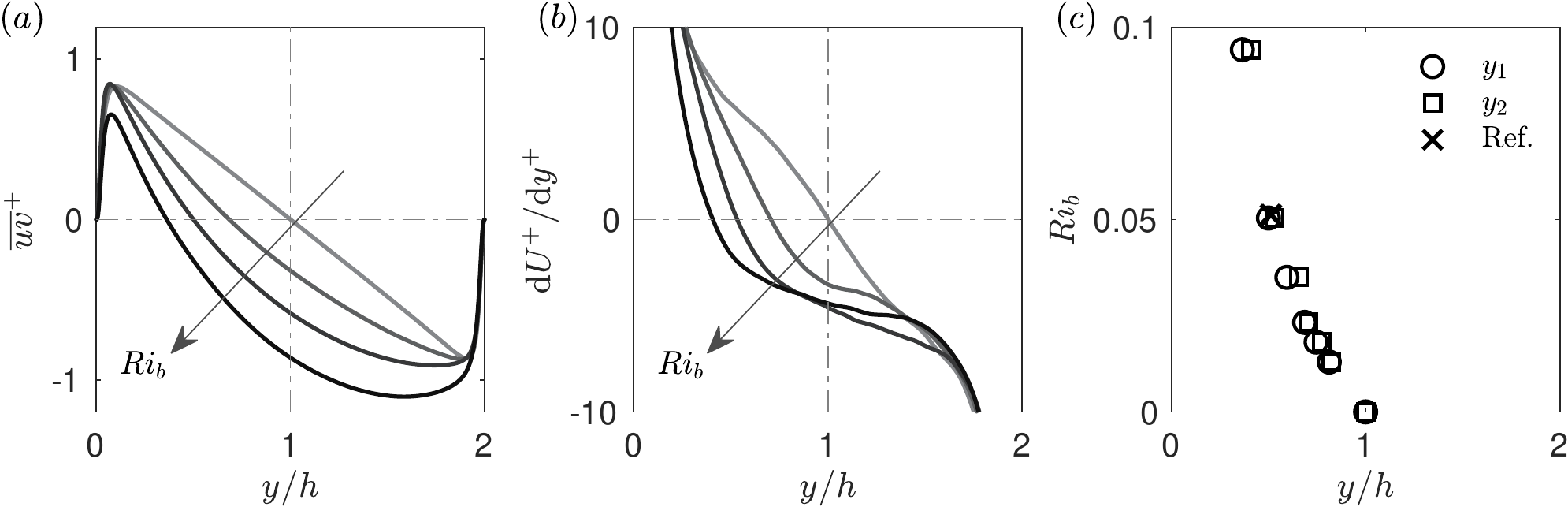}
			\caption{Position of the zero point in VMC: $(a)$ $y_1$ is the zero point of $\overline{uv}$ , and $(b)$   $y_2$ is the zero point of ${{\mathrm{d} U}/{\mathrm{d} y}}$ in the whole domain (hot wall $y/h=0$, cold wall $y/h=2$). The greyscale from light to heavy that varies with the increase in $Ri_b$, which is $0, 0.023, 0.050, 0.094$, is based on \cite{sutherland2015law}; see Table~\ref{tab:database} for detailed DNS cases.   $(c)$ is the variation in  $y_1$ and $y_2$ with the change in $Ri_b$, where the \emph{Ref. $\times$} is a verification case in \cite{kasagi1997direct} at $Ri_b =0.051$ (according to Figure~10, for the stress balance, Case 3f, $Re_\tau = 150$ and $Ra = 6.8\times 10^5$).}
			\label{fig:mxc_loc}
		\end{center}
	\end{figure*}
	
	For vertical mixed convection, Fig.~\ref{fig:mxc_loc} $(a)$ and $(b)$ shows that when $Ri_b = 0$ (forced convection), the profiles of  $\overline{uv}$ and ${{\mathrm{d} U}/{\mathrm{d} y}}$ have symmetry and the zero points stay at the centreline ($y/h=1$). 
	With an increase in $Ri_b$, both profiles gradually shift to the hotter wall side.  Fig.~\ref{fig:mxc_loc} $(c)$ illustrates the zero points $y_1$ and $y_2$ are almost at the same position until $Ri_b \simeq 0.10$. {It is still not clear whether $y_1$ and $y_2$ are mathematically the same.}
	Nevertheless, the mismatch between $y_1$ and $y_2$ is negligible.  
	Thus, the LEVM is approximately valid in the whole domain for $0 \le Ri_b < 0.10$. 
	The resulting {adjustment} region is illustrated in Fig.~{\ref{fig:mxc_prt}} for $Pr_{t}$ and Fig.~{\ref{fig:mx_eddy}} for $\nu^{dns}_t$. It is worth noting that
	$Pr_t$ is larger than unity within the viscous sublayer ($y^+ < 10$), which is quite different with respect to the buoyant horizontal channel \citep{pirozzoli2017mixed}. Meanwhile, the discrepancy of $Pr_{t}$ between the hotter and colder wall at the same $y^+$  becomes larger  with the increase in $Ri_b$. 
	To summarize, 
	the most distinctive feature is the break of symmetry for the mean and second moment statistics for VMC compared with the pure shear or buoyancy-driven vertical flow and horizontal channel \citep{garcia2011turbulence, pirozzoli2017mixed}.  
	Moreover,
	The asymmetry causes several modelling issues, such as the implementation of wall distance $y_w$ or $y^+$ for the low-$Re$ approach. 
	\begin{figure*}[!ht]
		\begin{center}
			\includegraphics[width=1.\textwidth]{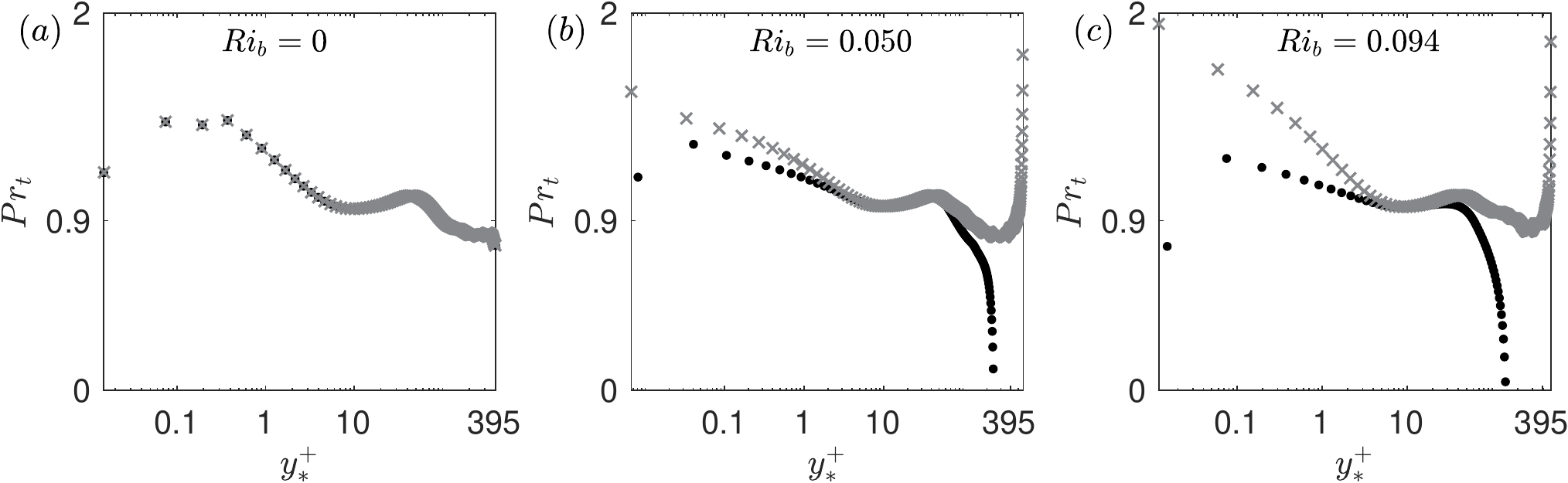}
			\caption{Turbulent Prandtl number ($Pr_{t}$)  for VMC, where $(a)$ $Ri_b =0$; $(b)$ $Ri_b =0.050$; $(c)$ $Ri_b = 0.094$, where $\newmoon$ indicates being near the hotter wall; {\color{gray} $\times$} indicates being near the colder wall.  The wall unit, $y^+_* = y^+_h=y_w U_{\tau, h}/\nu$ at the hotter wall and $y^+_* = y^+_c=y_w U_{\tau, c}/\nu$ at the colder wall. 
			}
			\label{fig:mxc_prt}
		\end{center}
	\end{figure*}
	\section{Modelling Methodologies}
	\label{sec:gep}
	In this section, we present the modelling framework via GEP with \emph{frozen} and \emph{CFD-driven} training.
	Then, the RANS-based approximation method for recovering DNS-based input quantities is introduced.  
	
	\subsection{Training framework}
	For modelling scalar flux, the goal is to find a mathematical representation of $\overline{u_i \theta}$.  Based on  dimensional arguments, \citet{shih1993remarks} showed that $\overline{u_i \theta}=f\left(U_{i,j}, \Theta_{,i}, k, \varepsilon, \overline{\theta^2}, \varepsilon_\theta\right)$, where $ \overline{\theta^2}$ denotes the temperature variance and $\varepsilon_\theta$ denotes the dissipation rate in a thermal field. This can be further simplified by assuming a sole time scale ($\overline{\theta^2}/\varepsilon_\theta \simeq k/\varepsilon$), which means that the thermal to mechanical time ratio, $\mathcal{R} = \overline{\theta^2} \varepsilon /(k \varepsilon_\theta)$  is treated as a  near unity constant \citep{dol1999dns}. Hence, we obtain $\overline{u_i \theta}=f\left(U_{i,j}, \Theta_{,i}, k, \varepsilon\right)$. 
	In light of Galilean invariance and nondimensionalization, a dimensionless velocity invariant $I$ and a  dimensionless temperature invariant  $J$ \citep{weatheritt2020data} are used to construct the target scalar flux models, 
	\begin{linenomath}
		\begin{align}
			& I=\left(c_\mu \frac{k}{\varepsilon}\right)^2 S_{ij}S_{ji}=\frac{1}{2}\left(c_\mu \frac{k}{\varepsilon}  \frac{\mathrm{d} U}{\mathrm{d}y} \right)^2,\\
			& J=\left(c_\mu \frac{k^{1.5}}{\varepsilon}  \frac{\mathrm{d} \Theta}{\mathrm{d}y} \right)^2,
		\end{align}
	\end{linenomath}
	where the mean stain rate tensor $S_{ij}=\frac{1}{2} \left(U_{i,j}+U_{j,i}\right)$, $c_\mu=0.09$.
	We adopt a variable turbulent Prandtl number $Pr_t$ with the reciprocal form $f(I, J) = {Pr^{-1}_t}$ and calculate the turbulent thermal diffusivity $\alpha_t$, 
	\begin{linenomath}
		\begin{equation}
			\alpha_{t} = \frac{\nu_t}{Pr_{t}}= {f(I,J) \nu_t}.
		\end{equation} 
	\end{linenomath}
	Note that the commonly used SGDH adopts  a constant turbulent Prandtl number $Pr_t=0.80\sim1.10$ (equivalent to $1/f(I,J)$ in the models proposed here) for air.
	
	\subsubsection{Frozen training}
	The machine learning procedure employs an in-house symbolic regression tool based on GEP, which was initially developed and tested  for Reynolds stress closures by \cite{weatheritt2016novel} and was recently used in scalar flux modelling \citep{weatheritt2020data}.  In general, we treat the wall-normal heat flux $\overline{v\theta}$ as a target term and regress by the constraint of cost function $J(\overline{v\theta})$; see Eq.\ref{eq:jvc}, where  a square root error is calculated along the wall-normal direction of $\varphi={\overline{v\theta}}$, with superscript \emph{dns} representing data from direct numerical simulation, and \text{\emph{gep} the value from simulation by GEP models}). 
	\begin{linenomath}
		\begin{equation} \label{eq:jvc}
			J(\varphi) =\int_{\mathrm{d}y} \left( \varphi^{dns}-\varphi^{gep}\right)^2 \mathrm{d}y
		\end{equation}
	\end{linenomath}
	This approach is called \emph{frozen} training \citep{zhao2020rans} as we are trying to optimize a closure against a fixed high-fidelity database. The detailed procedure can be found  in Algorithm 1. Moreover, before running GEP, there are two preprocessing steps. One possible issue is the lack of a DNS-based dissipation rate $\varepsilon$ suitable for the modelling. We overcome this issue by solving the transport equation of $\varepsilon$ (see Eq.~\ref{eq:eps}) with $ c_{1 \varepsilon}=1.44$,  $ c_{2 \varepsilon}=1.92$,  $ \sigma_{\varepsilon}=1.3$, where all the other quantities, such as $\overline{v\theta}, U, \Theta, k$, are extracted from DNS
	\begin{linenomath}
		\begin{equation}
			-\frac{\varepsilon}{k} \left[ c_{1\varepsilon} \left(\mathcal{P}+ \mathcal{G}\right)  - c_{2\epsilon} \varepsilon \right]= \frac{\mathrm{d}}{\mathrm{d} y} \left[ \left( \nu + \frac{\nu_t}{\sigma_\varepsilon}  \right) \frac{\mathrm{d}\varepsilon}{\mathrm{d} y} \right]
			\label{eq:eps}
		\end{equation}
	\end{linenomath}
	and $\mathcal{P} =-\overline{uv}\:{\mathrm{d} U}/{\mathrm{d} y}, \mathcal{G} = g \beta \overline{u\theta} $.
	Another step is the way to approximate $\nu_t$ (the infinity issue discussed  in \textsection \ref{sec:inf}), which is delineated in \textsection \ref{sec:nut}.
	
	\begin{figure}[!ht]
		\removelatexerror
		\begin{algorithm}[H]
			\caption{\emph{frozen} training}
			Get  $\overline{v\theta}, U, \Theta, k$ from DNS data, \\
			\If{do not have suitable DNS-based $\varepsilon$}{
				Solve transport equation of $\varepsilon$ (Eq.~\ref{eq:eps}) with DNS-based $\overline{v\theta}, U, \Theta, k$ as input\\} 
			Calculate positive smoothed $\nu^{mod}_t = f_\mu c_\mu k^2/\varepsilon$ for VMC, Eq.~\ref{equ_eddy_smooth} for VNC \\
			\For{\textrm{each generation step} $i=1,2,...,N$}{ 
				randomly generate population in the environment \\
				\For{\textrm{each population step} $j=1,2,...,M$}{
					genetic evolution to find the best candidate models based on the minimum fitness $J(\overline{v\theta})$ for the $i^{th}$ generation 
				}
			}
			${\text{Solve RANS scalar equation with new model to obtain } \Theta, Nu}$ \\
			\label{alg:frozen}
		\end{algorithm}
	\end{figure}
	
	\subsubsection{CFD-driven training}
	The resulting data-driven models via \emph{frozen} training can improve the performance of $\overline{v\theta}$, but sometimes the improvement fails to be shown in the mean flow field $\Theta$ and Nusselt number $Nu$. To find better turbulent heat flux models, we implement a loop algorithm that integrates GEP and a RANS solver, referred to as \emph{CFD-driven} training \citep{zhao2020rans} (see Algorithm 2). The major modification compared with \emph{frozen} training is solving the RANS scalar equation for each candidate model and then obtaining the RANS-based $\Theta$, $\mathrm{d}\Theta/\mathrm{d}y$ and $Nu$ to calculate the cost function (see Eq.\ref{eq:j}) in terms of quantities of interest. For example, the cost function can be the error of mean temperature $J(\Theta)$ (Eq.\ref{eq:jvc}, where $\varphi=\Theta$). We can also use the absolute error of $Nu$ ($J(Nu)$ in Eq.\ref{eq:jnu}), or the combination of errors, such as $J(cNu)$ in Eq.~\ref{eq:jcnu} and $J(cdc)$ in Eq.~\ref{eq:jcdc}.
	\begin{figure}[!ht]
		\removelatexerror
		\begin{algorithm}[H]
			\caption{\emph{CFD-driven} training}
			Get $\overline{v\theta}, U, \Theta, k$ from DNS data, calculate $\varepsilon, \nu_t$ (same as \emph{frozen} training)\\
			\For{\textrm{each generation step} $i=1,2,...,N$}{ 
				randomly generate population in the environment \\
				\For{\textrm{each population step} $j=1,2,...,M$}{
					feed $f(I,J)$ into RANS scalar equation for $\alpha_t$  \\
					solve the RANS scalar equation for each candiate model for the $i^{th}$ generation and obtain $\Theta, Nu$ \\
					find best candidate models constrained by customized cost ${J(\varphi) \text{ based on } \Theta \text{ or }\mathrm{d}\Theta/\mathrm{d}y}$ for the $i^{th}$ generation
				}
			}
			\label{alg:loop}
		\end{algorithm}
	\end{figure}
	\begin{linenomath}
		\begin{subeqnarray} \label{eq:j}
			J(Nu) = \frac{|Nu^{dns} - Nu^{gep}|}{Nu^{dns}} \times 100 \%, \slabel{eq:jnu} \\
			J(cNu) = J(\Theta) + \lambda J(Nu),   \slabel{eq:jcnu} \\
			J(cdc) = J(\Theta) + \lambda J(\mathrm{d}\Theta/\mathrm{d}y).   \slabel{eq:jcdc} 
		\end{subeqnarray}
	\end{linenomath}
	\subsection{Data preparation of DNS-based input quantities}
	\label{sec:nut}
	In \textsection \ref{sec:inf}, a close inspection of the flow features at different regions allows us to identify some issues with the linear gradient-based assumption and the validity of LEVM, especially the near-wall region for VNC.
	In practice, as the mean velocity gradient $\mathrm{d} U/\mathrm{d} y \to 0$ at $y_2$, that is, $\nu^{dns}_t$ could be a non-physical value in the vicinity of the wall region {as it tends to  $ \nu^{dns}_t \to \pm \infty$}
	\begin{linenomath}
		\begin{equation} \label{equ:eddy}
			\nu^{dns}_t = \frac{-\overline{uv}}{{\mathrm{d} U}/{\mathrm{d}  y}}
		\end{equation}
	\end{linenomath}
	Here, by assuming a smooth and positive $\nu_{t}$  \citep{xu1998new},  for VNC, we devise a limiter function (see Eq.~\ref{equ_eddy_smooth}) under the condition $\gamma$, 
	where $\gamma$ is $|{\mathrm{d} U}/{\mathrm{d}  y}|  \le1.2 \:\bigcap\:y_w/H <0.12$ to remove the singularity for the region of the infinity anomaly. The empirical constants in  Eq.{\ref{equ_eddy_smooth}} are obtained according to DNS data, which can be seen in Fig.~{\ref{fig:vnc_loc}} $(b)$. More specifically, $y_w/H < 0.12$ gives the upper bound of {the near-wall region for the smoothing operation} and $|{\mathrm{d} U}/{\mathrm{d}  y}|  \le1.2$ provides a good estimate for the infinity anomaly region for all $Ra$ from $10^5$ to $10^9$.
	Fig.~\ref{fig:vnc_eddy} illustrates the performance of Eq.~\ref{equ_eddy_smooth} at $Ra = 5.4\times 10^5$ and $Ra = 1.0\times 10^8$. It is clear that this limiter function can successfully remove the singularity and smoothly link the near-wall laminar and the bulk turbulent regions. 
	\begin{linenomath}
		\begin{equation}  \label{equ_eddy_smooth}
			\nu^{mod}_t=
			\begin{cases} 
				\frac{|\overline{u v}|} {{max(|\mathrm{d} U}/{\mathrm{d}  y|,1.2)}} & \quad \text{if } \text{condition } \gamma  \\
				\frac{|\overline{u v}|} {{ | \mathrm{d} U}/{\mathrm{d}  y|}}  & \quad \text{else} 
			\end{cases}
		\end{equation}
	\end{linenomath}
	
	\begin{figure*}[!ht]
		\begin{center}
			\includegraphics[width=1.0\textwidth]{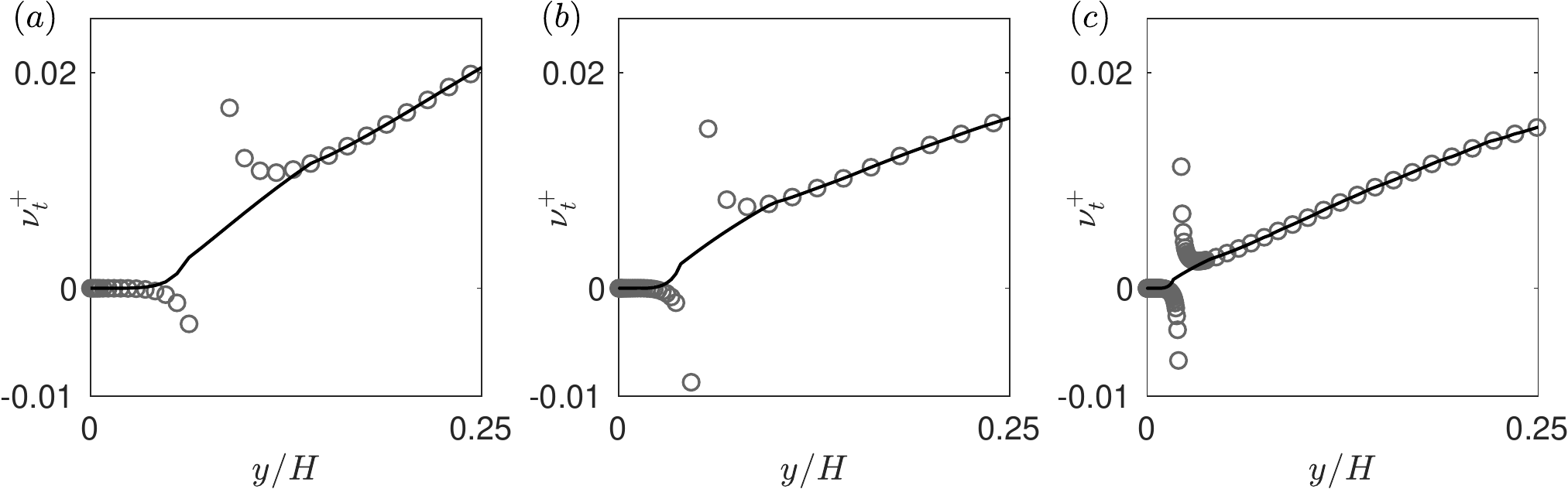}
			\caption{The smoothed turbulent eddy viscosity $\nu^+_t = \nu_t/(H U_f)$ for VNC: $(a)$, $Ra = 5.4\times 10^5$ ;  $(b)$, $Ra = 2.0\times 10^7$; $(c)$, $Ra = 1.0\times 10^9$, where $\fullmoon$ indicates DNS-based $\nu_t^{dns}$ (see Eq.~\ref{equ:eddy});  $\solid$ indicates $\nu_t^{mod}$  based on Eq.~\ref{equ_eddy_smooth}.}
			\label{fig:vnc_eddy}
		\end{center}
	\end{figure*}
	
	A treatment of the eddy viscosity is also needed in VMC. Unlike the anti-symmetric mean profile in VNC, the asymmetry increases the complexity of identifying the  position of $y_2$ (see Fig.~\ref{fig:mxc_loc}) at different $Ri_b$.
	Hence, we use a damping function $f_\mu$  based on the low Reynolds number modelling approach instead of the limiter function in VNC. The damping function $f_\mu$ \citep{myong1990new} is 
	\begin{linenomath}
		\begin{equation} 
			f_\mu = 1 + 3.45/\sqrt{Re_t}\left[1-exp(-y^+/70)\right]^2,
			\label{eq:fmuR}
		\end{equation} 
	\end{linenomath}
	where local Reynolds number $Re_t = k^2/(\nu \varepsilon)$, the viscous length scale $y^+ = y_w U_\tau/\nu$. 
	Eq. \ref{eq:fmuR} is compared to DNS-based  $f^{dns}_\mu$   for the $Ri00$ case (forced convection ) in Fig.~\ref{fig:fc_eddy} ($a$), in which we can see that the near-wall  (within $y^+<10$) prediction is fairly good. 
	\begin{figure*}[!ht]
		\begin{center}
			\includegraphics[width=1.\textwidth]{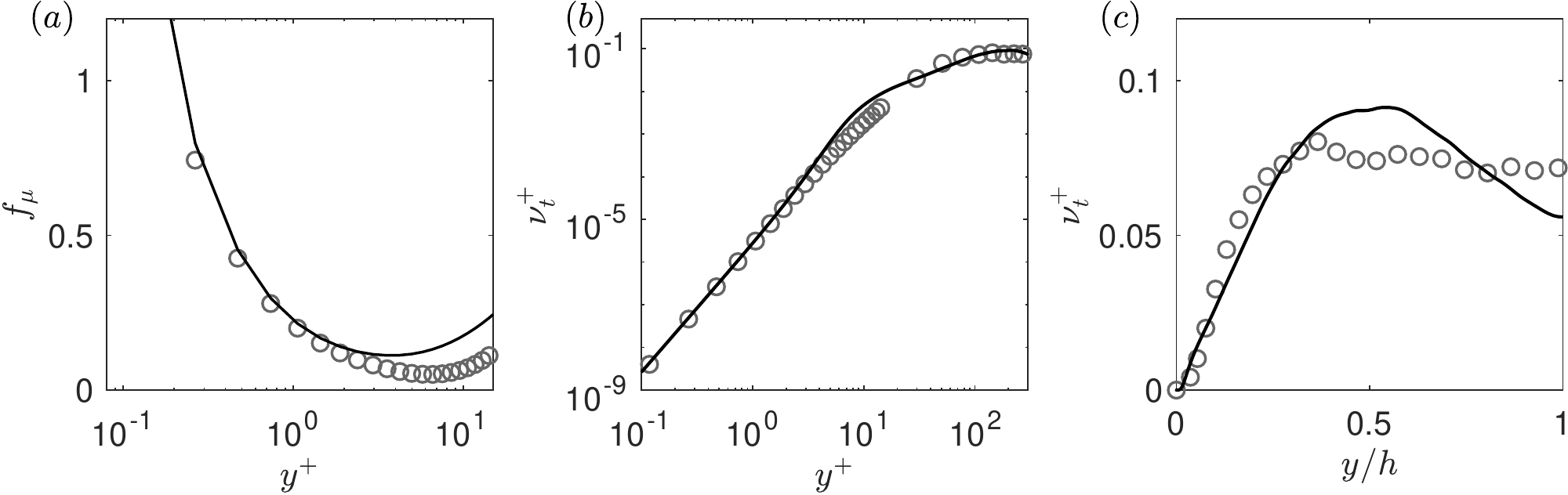}
			\caption{The turbulent eddy viscosity $\nu^+_t = \nu_t/U_\tau h$ for forced convection, ($a$) distribution of the damping function,  Eq. \ref{eq:fmuR}, $\solid$;  $f^{dns}_\mu = \frac{1}{c_\mu} \frac{-\overline{uv}} {{\mathrm{d} U}/{\mathrm{d}  y}} \frac{\varepsilon}{k^2}$, $\fullmoon$. The approximated $\nu^{mod}_t = f_\mu c_\mu k^2/\varepsilon$ is $\solid$; DNS-based $\nu^{dns}_t$ is $\fullmoon$ in  wall units in ($b$) and in global coordinates in ($c$).}
			\label{fig:fc_eddy}
		\end{center}
	\end{figure*}
	Hence, $\nu^{mod}_t = f_\mu c_\mu k^2/\varepsilon$ can be estimated quite well (Fig.~\ref{fig:fc_eddy}).
	The damping function is further tested in the VMC case. Fig.~\ref{fig:mx_eddy} shows that $f_\mu$ finds the best approximation with respect to several other damping functions \citep[see][for a review on the low Reynolds number modelling approach]{rodi1993low}. Note that the friction velocity at the hot and cold wall is different. Hence, Figs.~\ref{fig:mx_eddy} ($b$) and ($c$)  give the comparison at both walls in wall units. 
	\begin{figure*}[!ht]
		\begin{center}
			\includegraphics[width=1.\textwidth]{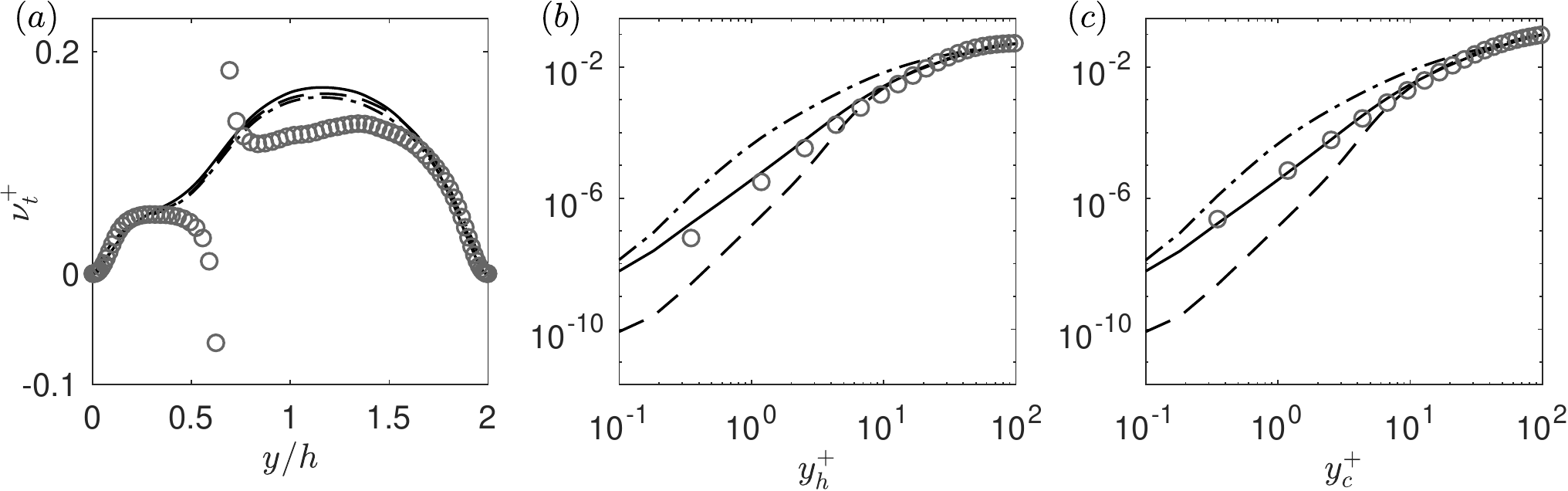}
			\caption{The turbulent eddy viscosity $\nu^+_t = \nu_t/U_\tau h$ for VMC at $Ri_b = 0.050$, approximated by Eq. \ref{eq:fmuR}, where $\nu^{mod}_t = f_\mu c_\mu k^2/\varepsilon$ is indicated by $\solid$ and DNS-based $\nu^{dns}_t$ is indicated by $\fullmoon$. The other damping functions are the standard $k-\varepsilon$ model where $f_\mu = 1$, $\dashed$; and the Lam-Bremhorst model where $f_\mu = \left[ 1-exp(-0.0165 \sqrt{k} y_w/\nu)\right]^2 (1+20.5/Re_t)$, $\chaindot$. ($a$) in global coordinates; ($b$) in viscous wall units at the hotter wall $y^+_h = y_w U_{\tau, h}/\nu$; ($c$) in viscous wall units at the colder wall $y^+_c = y_w U_{\tau, c}/\nu$.}
			\label{fig:mx_eddy}
		\end{center}
	\end{figure*}
	\section{Results}
	\label{sec:res}
	In this section, the machine learnt models are first presented. Then, we appraise the performance of the models resulting from various training datasets and approaches by comparisons with the DNS database.  Last, an \emph{a priori} test on the turbulent Prandtl number and an \emph{a posteriori} assessment for quantities of interest are shown. 
	\subsection{Machine learnt models}
	The GEP training approaches (Algorithms 1 and 2) are applied the dataset in Table~\ref{tab:database}. Here, we delineate the detailed information of a machine learning case for VNC at $Ra=2.0\times10^7$. Fig.~\ref{fig:trainig_proc} shows the normalized error metric (in this case, $J(\overline{v \theta})$ via \emph{frozen} training) in the training process. The population mean error dramatically decays until the 100th generation during the GEP evolution process.  After the 100th generation, the population mean error fluctuates around a fixed value, and the population minimum stays at a relatively small value. Thus,  the 100th generation candidate model is chosen, which is 
	\begin{linenomath}
		\begin{equation} \label{eq:mod_eg}
			f(I, J)=\underbrace{1.116}_{const.} + \underbrace{(0.205I - 12 J) }_{invariants}.
		\end{equation}
	\end{linenomath}
	
	\begin{figure}[!ht]
		\begin{center}
			\includegraphics[width=0.45\textwidth]{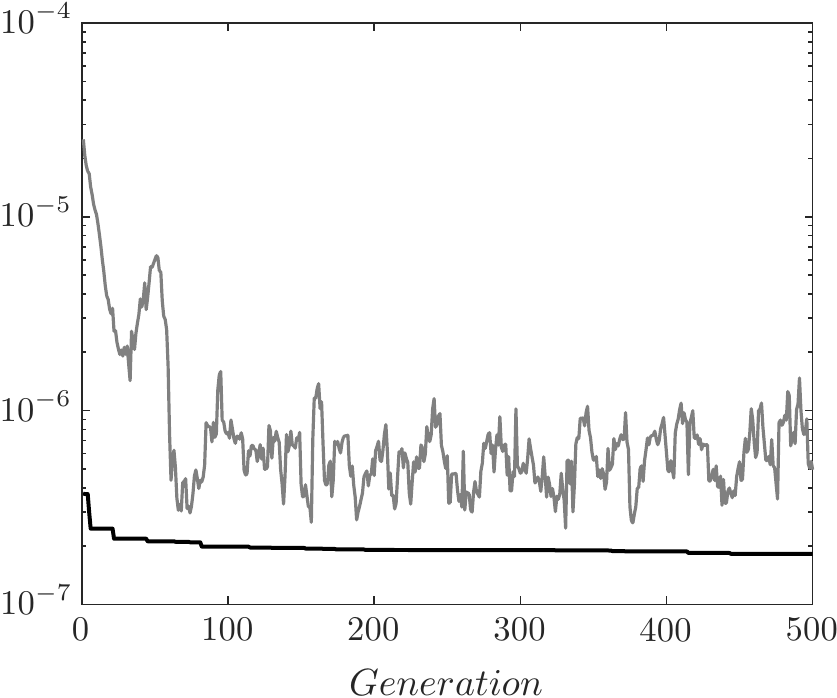}
			\caption{Evolutionary algorithm performance: {\color{gray}\solid}, (grey solid line) the population mean; {\color{black}\solid}, (black solid  line) the population minimum.}
			\label{fig:trainig_proc}
		\end{center}
	\end{figure}
	
	For \emph{CFD-driven} machine learning, the computer costs are augmented by solving a RANS scalar equation for the candidate models in each generation.
	For a personal computer (4 cores, Intel core i7--7500U), {the training procedure, involving around $2\times 10^4$ RANS calculations (100 generations and 200 population size), takes approximately 27 hours. (Each RANS calculation takes around 20 seconds on one core to converge with the baseline results as initial condition)}.  
	Besides, the resulting models are more compact than the \emph{frozen} approach, as indicated by \cite{zhao2020rans}, {and have the form of $f(I, J)=0.969 + 2 J$,  for instance, in case of \textit{ncgrad}.}

	For RANS model development in turbulent wall-bounded flow, a long-standing struggle is to satisfy the wall asymptotic behaviour.
	Due to the no-slip condition for velocity terms,  the {isothermal} condition at both walls, and the continuity equation,  we can  derive that $v \propto{\mathcal{O}(y^2)}$, $u, w, \theta \propto { \mathcal{O}(y^1)}$, where $\propto{\mathcal{O}(y^n)}$ indicates a quantity is proportional to the nth order of the wall-normal coordinate. 
	Then, it directly suggests  $\overline{v \theta}\propto { \mathcal{O}(y^3)}$, ${\frac{\mathrm{d} \Theta}{\mathrm{d} y}} \propto{\mathcal{O}(y^0)}$. 
	For our model target, $\nu_t \propto { \mathcal{O}(y^3)}, \alpha_t \propto { \mathcal{O}(y^3)}, Pr_t \propto { \mathcal{O}(y^0)}$; hence, $f(I,J) \propto \mathcal{O}(y^0)$.  
	In Eq.~\ref{eq:mod_eg}, Table \ref{tab:vnc_mod} and Table \ref{tab:mxc_mod}, the GEP resulting models always have a constant term {and thus satisfy $\mathcal{O}(y^0)$.}
	The other terms consist of invariants
	$I  \propto { \mathcal{O}(y^4)}$ and $J \propto { \mathcal{O}(y^6)}$, which means the $I$ and $J$ terms do not affect the near-wall asymptotic behaviour. 
	Overall, comparing $f(I,J) \propto \mathcal{O}(y^0)$, it indicates that the GEP approach can ensure the correct wall-limiting behaviour.  
	All the  \emph{a priori}  and \emph{a posteriori} assessments for these models are described in the following text. 
	
	
	\subsection{Sensitivity study on training datasets and training approaches}
	Following the same numerical treatment, we obtain a series of resulting GEP models  via \emph{frozen} and \emph{CFD-driven} training. Table~\ref{tab:vnc_mod} lists the models for the VNC cases, which are trained on $Ra=2.0\times10^7$. For the VMC cases, Table~\ref{tab:mxc_mod} shows the models that are trained on $Ri_b=0.05$.  In this section, the dependence on the training datasets and cost functions are presented to seek the best model in an \emph{a posteriori} sense. 
	
	\begin{table*}
		\centering
		\begin{tabular*}{.85\textwidth}{@{\extracolsep{\fill}}llll}
			\toprule
			Label & Training approach & Cost function & Heat flux models $f(I,J)$ \\
			\hline
			\emph{base} & -     & - &$1.111$ ($Pr_t=0.90$) \\
			\hline
			\emph{ncflux} & frozen &$J(\overline{v\theta})$   & $1.116 + 0.205I - 12 J$ \\
			\emph{ncmean} & CFD-driven & $J(\Theta)$   & $1. 195-I + J$\\
			\emph{ncnu} & CFD-driven & $J(Nu)$   & $1.031 - I$\\
			\emph{ncgrad} & CFD-driven &  $J(\mathrm{d}\Theta/\mathrm{d}y)$    & $0.969 + 2 J$\\
			\bottomrule
		\end{tabular*}
		\caption{GEP models for VNC based on $Ra=2.0\times10^7$ via GEP in form of $f(I,J)$.}  
		\label{tab:vnc_mod}
	\end{table*}
	\begin{table*}
		\centering
		\begin{tabular*}{.85\textwidth}{@{\extracolsep{\fill}}llll}
			\toprule
			Label & Training approach & Cost function & Heat flux models $f(I,J)$ \\
			\hline
			\emph{base} & -     & - &$1.111$ ($Pr_t=0.90$) \\
			\hline
			\emph{mcflux} & frozen &$J(\overline{v\theta})$   & $1.057-0.565 I-0.188 J$ \\
			\emph{mcmean} & CFD-driven & $J(\Theta)$   & $1.000 + 0.861 I (-0.215 + I - 0.5 J) $\\
			\emph{mcgrad} & CFD-driven &  $J(\mathrm{d}\Theta/\mathrm{d}y)$    & $1.099 + I (-1.180 - 0.900 J) $\\
			\hline
			\emph{mccnu}   & CFD-driven & $J(\Theta)+\lambda_1 J(Nu)$  & $0.970 - 0.305 I^2$ \\
			\emph{mccdc}   & CFD-driven  & $J(\Theta)+\lambda_2 J(\mathrm{d}\Theta/\mathrm{d}y)$   & $1.090 - I$ \\
			\bottomrule
		\end{tabular*}
		\caption{GEP models for VMC based on $Ri_b=0.050$ via GEP in the form of $f(I,J)$, The combination factor $\lambda$ is used to ensure the error metric at the same magnitude, where $\lambda_1 = 0.1$, $\lambda_2 = 20$. }  
		\label{tab:mxc_mod}
	\end{table*}
	
	\subsubsection{Dependence on training datasets}
	\label{sec:data_dep}
	Since the data-driven approach can depend on the training case, it is essential to cross-validate each model.  
	We adopt a holdout training and testing approach, where a $Ra$ or $Ri_b$ case are selected to train a model and other cases are used to test the performance of this model. The dependence on training datasets (or cases) is studied for both VNC and VMC.
	Fig.~\ref{fig:vnc_nu_error} and Fig.~\ref{fig:vnc_temp_error} lay out the error of $Nu$ and $\Theta$, respectively, with comparison of the \emph{a posteriori} performance across $Ra$ for VNC. 
	Note that we omit the other  GEP models for cross-validation.
	
	Fig.~\ref{fig:vnc_nu_error}($a$) presents the correlation map depicting the performance of trained models for $Nu$ in an \emph{a posteriori} sense via the frozen approach with cost function $J(\overline{v\theta})$.
	The legend is scaled by the absolute percentage error $J(Nu)$ of the baseline model, which is $28.7\%$ for the Ra80 case. 
	The maximum GEP-based $J(Nu)$ for all testing cases is $9.5\%$, which shows the significant improvement achieved by GEP training. 
	Moreover, the performance of the models trained on each $Ra$ is generally better than the performance for the other cases; see the diagonal component in Fig.~\ref{fig:vnc_nu_error}($a$), and interestingly, there are exceptional cases such as Ra90, where $J(Nu)$ is the largest for itself ($6.7\%$).
	Nevertheless, the best model via the \emph{frozen} approach resulting from the Ra63 case can reduce $J(Nu)$ to $3.5\%$ for all the VNC cases. 
	Similarly,  Fig.~\ref{fig:vnc_temp_error}($a$) shows the performance of trained models  with cost function $J(\overline{v\theta})$ for $\Theta$ in an \emph{a posteriori} sense.
	The legend is scaled by the square root error $J(\Theta)$ of the baseline model, which is $11.5 \times 10^{-3}$ for the Ra80 case. 
	In contrast to the universal {and significant} improvement seen for $J(Nu)$, the error reduction on $J(\Theta)$ is {relatively small} across different $Ra$ cases.   
	The reduction of predictive error is $6.1\%$ with respect to the maximum baseline error for Ra80 by the frozen trained models at Ra50.
	However, we can obtain a generalized GEP model.
	The best model via the \emph{frozen} approach results from the Ra73 case, which can reduce $J(\Theta)$ to $5.6 \times 10^{-3}$ for all the VNC cases, achieving approximately $50\%$ improvement. 
	In brief, 
	the machine learnt models are independent of  $Ra$ for VNC cases, and the GEP models via the \emph{frozen} approach generally perform better than baseline models, especially at the higher $Ra$ range.  
	\begin{figure*}[!ht]
		\begin{center}
			\includegraphics[width=1.0\textwidth]{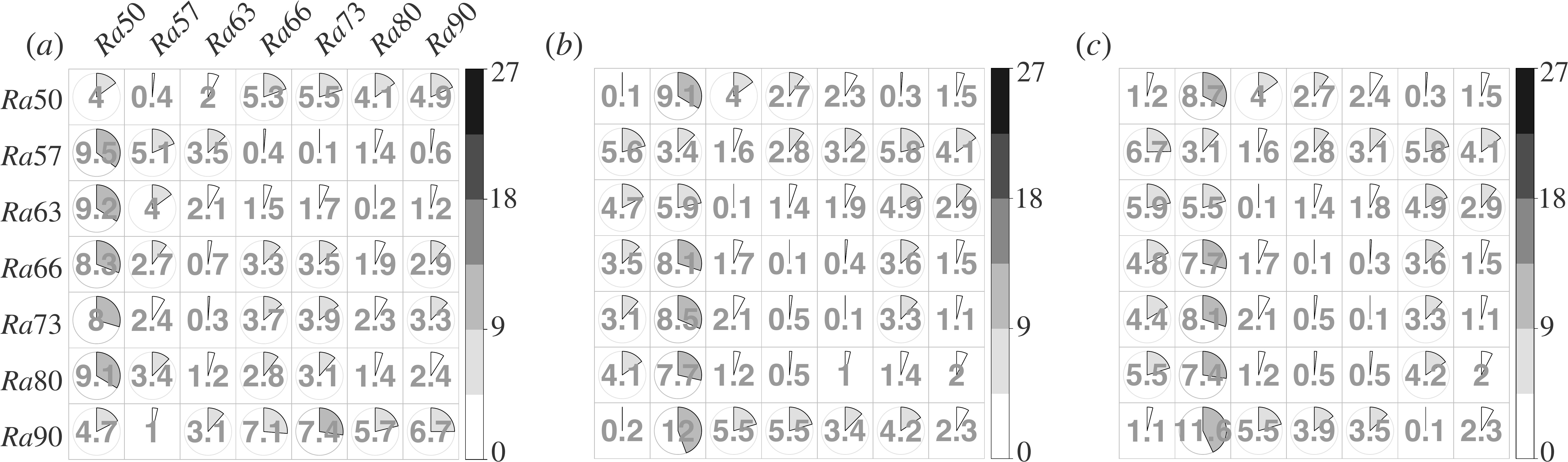}
			\caption{Error of $Nu$, premultiplied by 100,   ($J(Nu)$ see Eq.\ref{eq:jnu}) with various GEP models, $(a)$ \emph{ncflux}; $(b)$  \emph{ncnu}; $(c)$  \emph{ncgrad}.  The row label means training a GEP model at this $Ra$, the column means testing the performance of a model in an \emph{a posteriori} sense at this $Ra$. The color bar is based on the maximum $J(Nu)$ with baseline models, which occurs in the {Ra80} case ($J(Nu)=28.7\%$).}
			\label{fig:vnc_nu_error}
		\end{center}
	\end{figure*}
	\begin{figure*}[!ht]
		\begin{center}
			\includegraphics[width=1.0\textwidth]{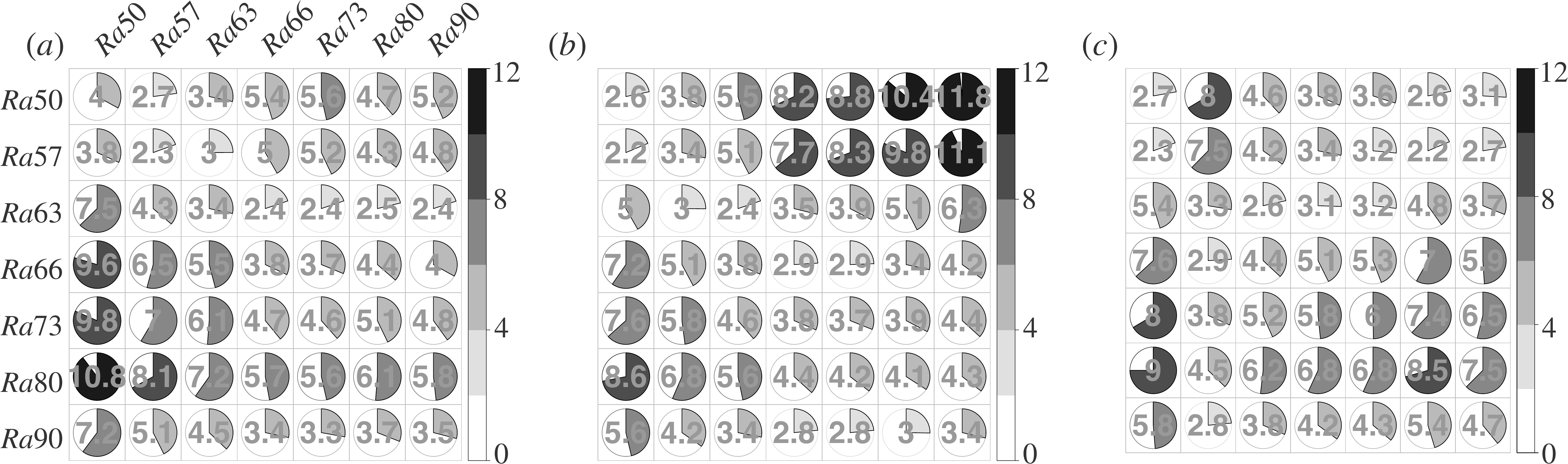}
			\caption{Error of $ \Theta$ ($J(\Theta)$, premultiplied by 1000,  see Eq.\ref{eq:jnu}) with various GEP models, $(a)$ \emph{ncflux}; $(b)$  \emph{ncmean}; $(c)$  \emph{ncgrad}. The row label means training a GEP model at this $Ra$, the column means testing the performance of a model in an \emph{a posteriori} sense at this $Ra$.  The color bar is based on the maximum $J(Nu)$ with baseline models, which occurs in the {Ra90} case.}
			\label{fig:vnc_temp_error}
		\end{center}
	\end{figure*}
	
	Fig.~\ref{fig:mx_data_error} illustrates the performance of GEP models trained  on Ri18, R50, and Ri94 cases via   \emph{frozen} training.  The baseline model perfectly captures $Nu$ for forced convection $Ri_b=0$. However, the prediction errors of both $Nu$ and $\overline{v \theta}$ linearly increase with the growth of the buoyancy effect (see Fig.~\ref{fig:mx_data_error} {$(a),(c)$}).
	Conversely, the error reductions of GEP models trained on the Ri94 case nearly linearly increase with the decrease in the buoyancy factor. 
	The resulting models trained on Ra50 significantly reduce the error of $Nu$ to $5\%$.
	Moreover, all the GEP models can reduce the prediction error of $\overline{v\theta}$.
	Surprisingly, 
	for the mean temperature, the baseline model is better than all the GEP modes trained on different $Ri_b$ cases via \emph{frozen} training except the largest $Ri_b$. 
	To summarize, 
	the GEP models depend on different $Ri_b$ cases, where the middle range case Ri50 shows the best performance for $Nu$ if we regard the Ri00 case as an exception. 
	This result further suggests the limitation of the \emph{frozen} training approach.}

\begin{figure*}[!ht]
	\begin{center}
		\includegraphics[width=0.99\textwidth]{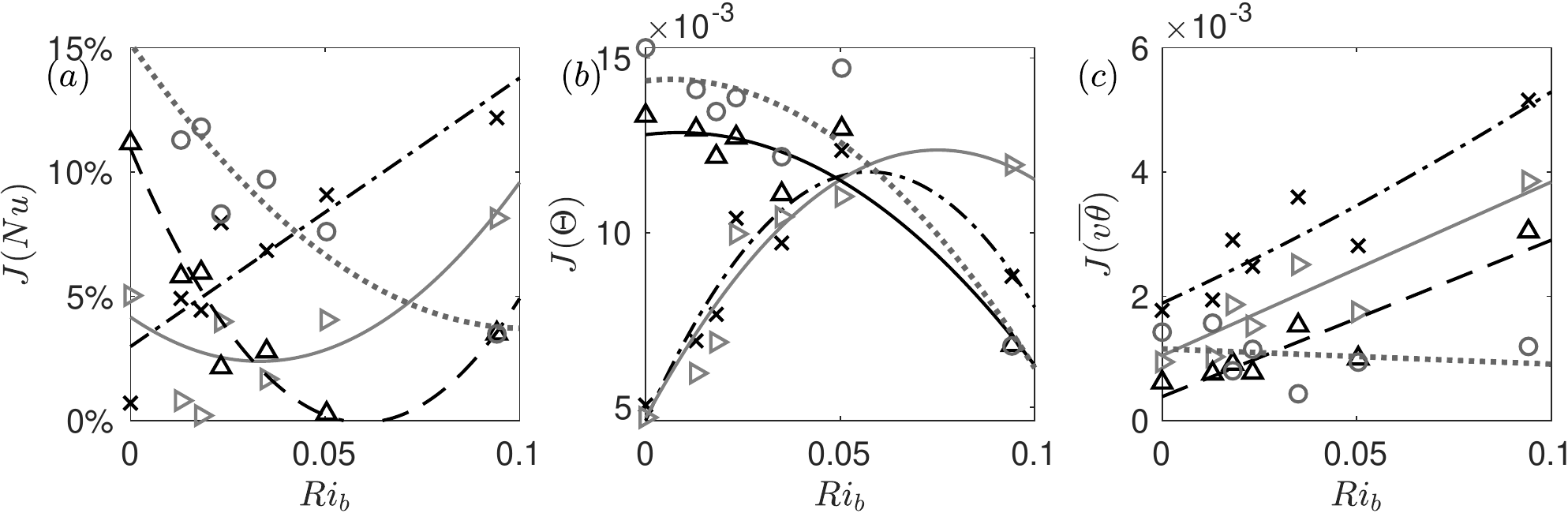}
		\caption{Error metric with the baseline or GEP models, $(a)$ error of Nusselt number $J(Nu)$ (see Eq.\ref{eq:jnu}); $(b)$ error of mean temperature (see Eq.\ref{eq:jvc}  where $\varphi = \Theta$);  $(c)$ error of wall-normal heat flux (see Eq.\ref{eq:jvc} where $\varphi =\overline{v\theta}$). At each $Ri_b$, $\times$, \emph{base}; $\triangleright$, trained on $Ri_b= 0.018$, $f(I,J) = 1.141 + I(0.387-0.387 J)-0.141J$,  $\triangle$, \emph{mcflux}, trained on $Ri_b= 0.050$; $\fullmoon$, trained on $Ri_b= 0.094$, $ f(I,J) = 0.966 - I - 2(-0.43 + I)(I - 0.089J)J  $,  \emph{mcgrad}. For each model, we plot a fitting curve to show the trend of error at different $Ri_b$, $\chaindot$, \emph{base}; $\dashed$, \emph{mcflux}, trained on Ri50; $\solid$,  trained based on {Ri18}, $\dashed$, trained based on {Ri94}. }
		\label{fig:mx_data_error}
	\end{center}
\end{figure*}

\subsubsection{Effects of training approaches}
The comparison of training approaches is the assessment on cost functions for the data-driven method. When we select $J({Nu})$ as the cost function, i.e. we use the \emph{CFD-driven} approach, the performance of GEP models on the prediction of $Nu$ is generally better than that
of using other cost functions. 
For instance, for VNC, 
Fig.~\ref{fig:vnc_nu_error}($b$) depicts the \emph{a posteriori} correlation error of  $Nu$ trained by $J(Nu)$ via the \emph{CFD-driven} approach and
shows the diagonal component, which means training and testing for the same case, is smaller than that of Fig.~\ref{fig:vnc_nu_error}($a$) (trained by $J(\overline{v\theta})$ {via the \textit{frozen} approach, i.e. without involving CFD while training}). 
This is also true for the \textit{a posteriori} predication of $J(\Theta)$, when the models are trained on $J(\Theta)$ (using the \textit{CFD-driven} approach), see Fig.~{\ref{fig:vnc_temp_error}}($b$), rather than when using training based on $J(\overline{v\theta})$ via the \textit{frozen} approach (Fig.~{\ref{fig:vnc_temp_error}}($a$)).
Accordingly, 
when we select $J(Nu)$ as the cost function, it can undermine the performance on $\Theta$ of the resulting GEP models and vice versa.
The maximum {\textit{a posteriori} error} of $J(Nu)$ trained by {cost function} $J(\Theta)$ is $18.3\%$, which is worse than training by $J(Nu)$.
However, 
the performance of GEP models constrained by $J(\mathrm{d} \Theta/\mathrm{d} y)$ (Fig.~\ref{fig:vnc_nu_error}($c$) and Fig.~\ref{fig:vnc_temp_error}($c$)) is slightly better than that of $J(\overline{v\theta})$.  
Overall, the model trained on Ra73 by $J(\mathrm{d} \Theta/\mathrm{d} y)$ (\emph{ncgrad} in Table~\ref{tab:vnc_mod}) seems to be the best model for VNC.               

\begin{figure*}[!ht]
	\begin{center}
		\includegraphics[width=0.99\textwidth]{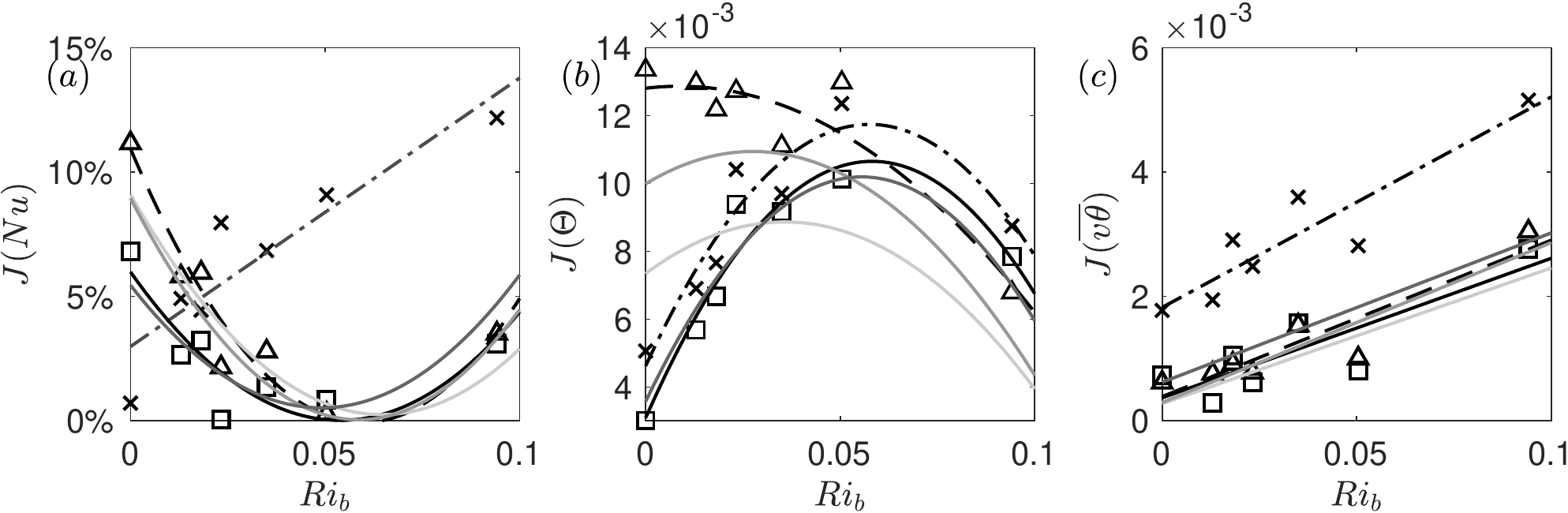}
		\caption{Error metric with baseline or GEP models, $(a)$ error of the Nusselt number $J(Nu)$ (see Eq.\ref{eq:jnu}); $(b)$ error of the mean temperature (see Eq.\ref{eq:jvc}  where $\varphi = \Theta$);  $(c)$ error of wall-normal heat flux (see Eq.\ref{eq:jvc} where $\varphi =\overline{v\theta}$). At each $Ri_b$, $\times$, \emph{base}; $\triangle$, \emph{mcflux}; $\square$, \emph{mcgrad}. For each model, we plot a fitting curve to show the trend at different $Ri_b$ cases, $\chaindot$, \emph{base}; $\dashed$, \emph{mcflux};  \emph{CFD-driven}, $\solid$, from light to heavy, \emph{mcmean}, \emph{nccnu}, \emph{mcdc}, \emph{mcgrad}, the heaviest (black) solid line,  \emph{mcgrad}. }
		\label{fig:mx_cost_error}
	\end{center}
\end{figure*}

For VMC, as stated in \textsection{\ref{sec:data_dep}}, the \emph{frozen} approach can reduce the error of $Nu$, but the performance on $\Theta$ is even worse than when using the baseline. 
Fig.~\ref{fig:mx_cost_error} shows the effect of various cost functions (see Table \ref{tab:mxc_mod}) based on Ra50. 
All the \emph{CFD-driven}-based results can achieve a better prediction than the frozen approach for both $Nu$ and $\overline{v\theta}$. 
Interestingly, the models developed with $J(\Theta)$ in the cases \emph{mcmean} and \emph{mccnu} (see Table~\ref{tab:mxc_mod}) perform better than the frozen approach but worse than the baseline model.
In contrast, they can reduce the error in $\Theta$ (see Fig.~\ref{fig:mx_cost_error} {$(b)$}) at the low $Ri_b$ regime for the cost function with the inclusion of $\mathrm{d}\Theta / \mathrm{d} y$ (cases \emph{mcgrad} and \emph{mcdc}).
Moreover, 
when we investigate the combination of quantities of interest, the \emph{mcdc} case is the best model for the VMC cases. 
To summarize, the influence of training approaches for VMC cases is more significant than that for VNC cases.  With the precondition of training on the middle range $Ri_b$ case, we finally find a machine learnt model via the \emph{CFD-driven} approach with the cost function $J(\Theta)+\lambda_2 J(\mathrm{d}\Theta/\mathrm{d}y)$ that performs well  for all the considered $Ri_b$ cases.
\subsection{\emph{A priori} test on the turbulent Prandtl number }
\begin{figure*}
	\begin{center}
		\includegraphics[width=1\textwidth]{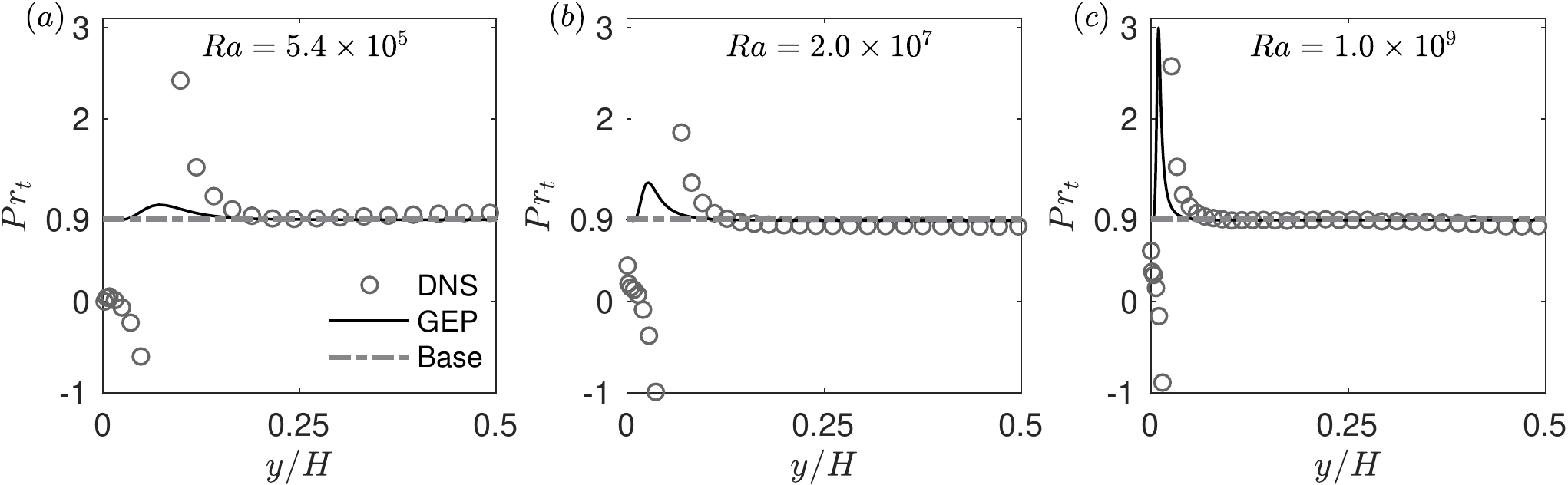}
		\caption{\emph{A priori} test on the turbulent Prandtl number for VNC based on the resulting models trained on $Ra=2.0\times10^7$.}
		\label{fig:nc_prt}
	\end{center}
\end{figure*}

The predicted turbulent Prandtl number for VNC is shown in Fig.~\ref{fig:nc_prt}, where the DNS results \citep{ng2015vertical} and baseline calculate with a constant $Pr_t = 0.90$ are included, with $Ra$ spanning four decades from $10^5$ to $10^9$.
The $Pr_{t}$ has the same feature.  
In the turbulent layer region ($y_2 < y_w< h$), $Pr^{dns}_{t}$ remains at a constant value, which stays in the range of $0.85\sim1.0$ across different $Ra$ numbers. 
Therefore, constant $Pr_{t}$ may turn out to be a good approximation as $Ra \to \infty$ and {the region of infinity anomaly} diminishes.
Conversely, it shows that the machine learnt model 
provides spatially varying $Pr_t$ in the near-wall region.  It is essential that  the resulting GEP model can identify the adjustment region ($y_1<y<y_2$) and bridge the infinity region of $Pr_{t}$ with a finite value, without any user intervention, due to the self-adapting feature of dimensionless frame invariants.   

\begin{figure*}[!ht]
	\begin{center}
		\includegraphics[width=1\textwidth]{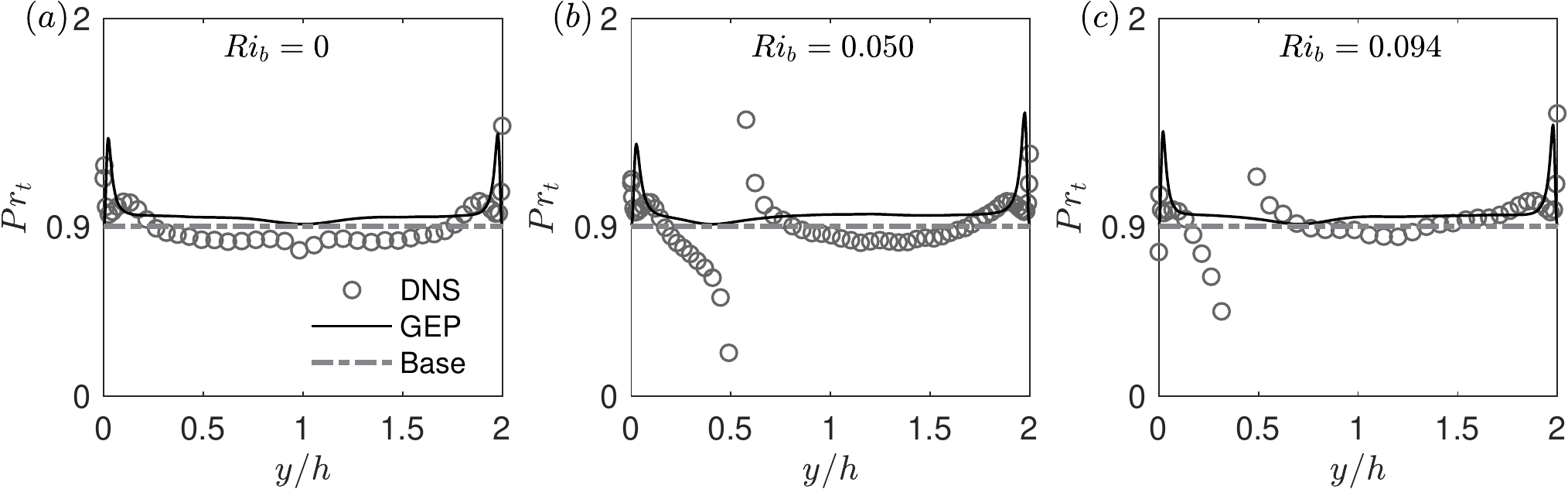}
		\caption{\emph{A priori} test on the turbulent Prandtl number for VMC based on the  models resulting from training on $Ri_b=0.050$.}
		\label{fig:mc_prt}
	\end{center}
\end{figure*}

In Fig.~\ref{fig:mc_prt}, the GEP generated $Pr_{t}$ for VMC (Table~\ref{tab:mxc_mod}, \emph{mcgrad}) is compared against the DNS-based $Pr_t$ \citep{sutherland2015law} and a constant $Pr_t = 0.90$ for different $Ri_b$. 
In the very near-wall region, $I$ and $J$ are near zero (owing to $k=0$  at the hotter ($y/h=0$) and colder ($y/h=2$) walls); hence,  $Pr_{t} = 1/f(I,J) = 0.91$, which is smaller than the DNS near-wall results. 
Away from each wall, the $Pr^{gep}_{t}$ quickly reaches a maximum (at approximately $1.4$) near $y^+ = 10$, while the maxima of $Pr^{dns}_{t}$ (see Fig.~\ref{fig:mxc_prt}) are close to each wall. However, they both decrease to $0.90$ before entering the infinity region (near $y_2$).  
Moreover, the agreement between GEP and DNS results is not as good compared with VNC case. One possible reason could be the discrepancy of the approximated eddy viscosity $\nu^{mod}_t$ {with DNS-based eddy viscosity} .
Nevertheless, the \emph{a priori} assessment shows GEP explicitly returns a variable $Pr_t$, and the infinity regions are approximated by values of nearly 0.90 across the different $Ri_b$ cases. 

\subsection{\emph{A posteriori} performance for quantities of interest}
\begin{figure*}[!ht]
	\centering
	\includegraphics[width=0.46\textwidth]{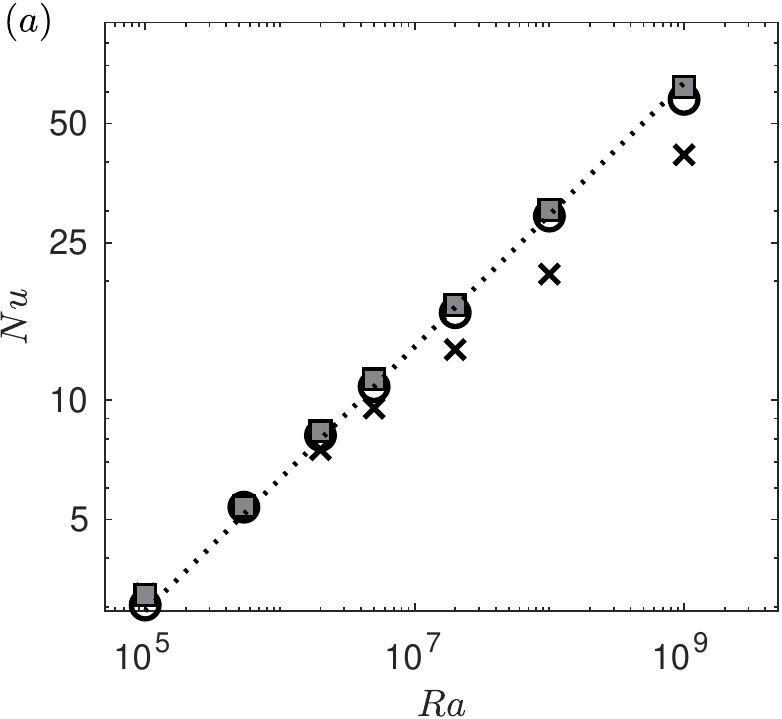}
	\hspace{2em}
	\includegraphics[width=0.46\textwidth]{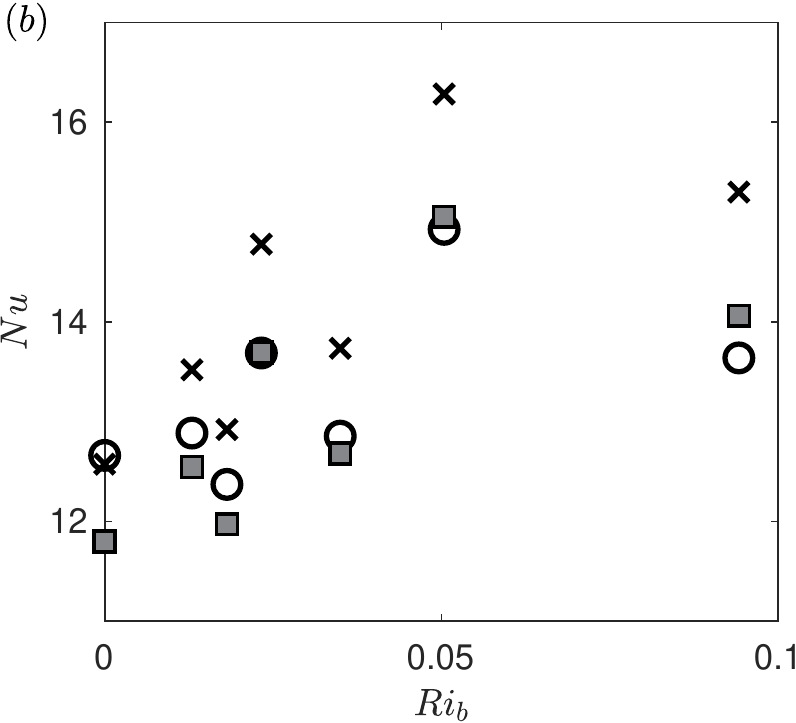}
	\caption{\emph{A posteriori} assessment of the Nusselt number, $\fullmoon$, DNS;   {\color{gray}$\blacksquare$}, GEP; $\times$, base; $(a)$ $Nu$ versus $Ra$ for VNC, on a log-log scale, {\color{gray}$\dotted$}, $Nu=0.071 (Ra Pr)^{1/3} $ by \cite{versteegh1999direct}; $(b)$ $Nu$ versus $Ri_b$ for VMC.}
	\label{fig:post_nu}
\end{figure*}

Fig.~\ref{fig:post_nu}$(a)$) shows the comparison of $Nu$ resulting from the baseline and GEP models with DNS data at different $Ra$ numbers. The baseline model significantly underpredicts the $Nu$, especially at higher $Ra$, with an absolute percentage error over $25\%$. Conversely, the GEP models can successfully predict the classical heat-transfer relationship $Nu\sim Ra^{1/3}$. 
Fig.~\ref{fig:post_nu}$(b)$ illustrates $Nu$ versus $Ri_b$ for VMC. 
Interestingly, as we showed earlier, the baseline can nearly perfectly predict $Nu$ in the forced convection case but considerably overpredicts $Nu$ with at least $10\%$ absolute percentage error for higher $Ri_b$ values. The performance of the GEP models reduces the error to less than $5\%$.
\begin{figure*}[!ht]
	\begin{center}
		\includegraphics[width=1\textwidth]{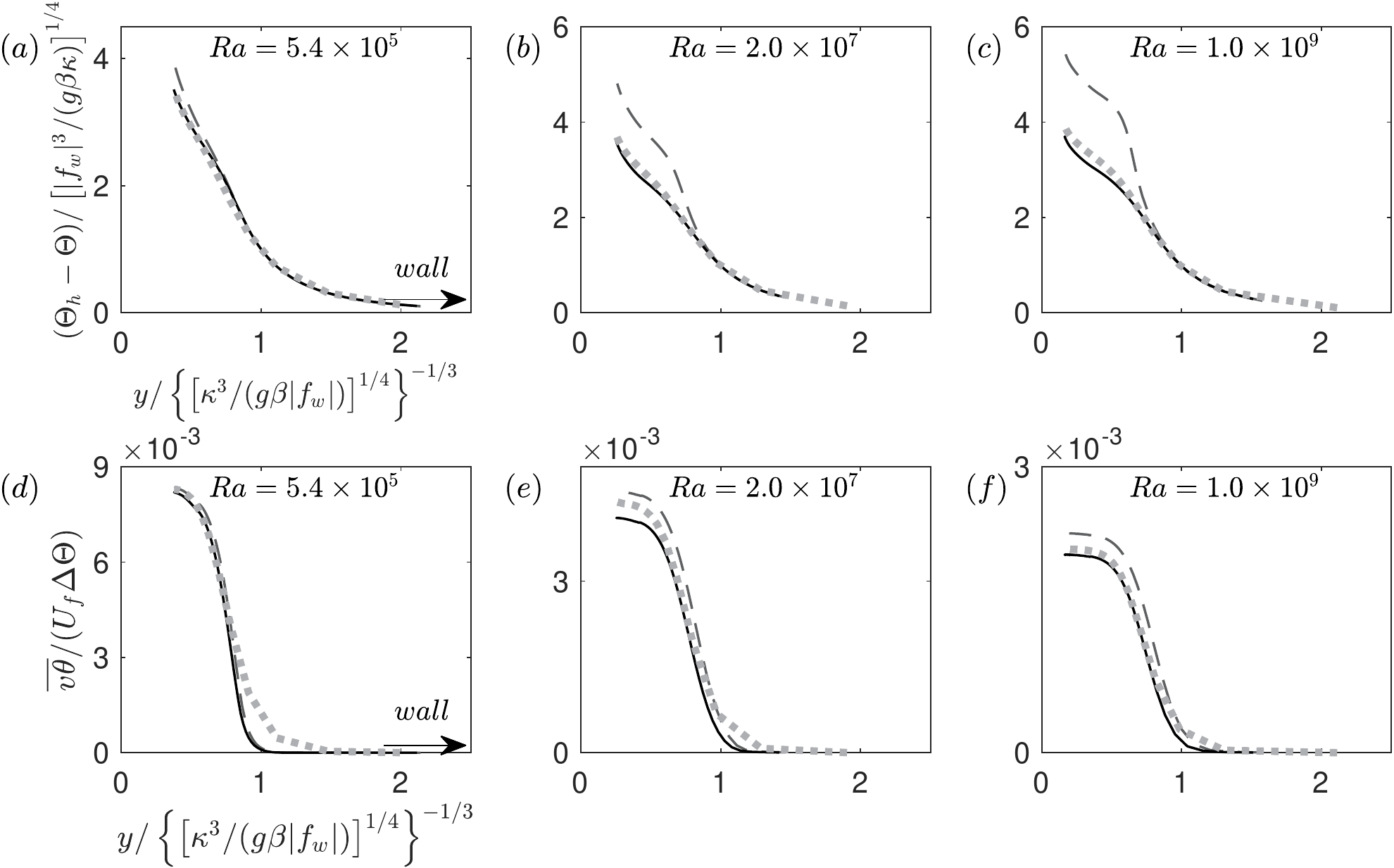}
		\caption{\emph{A posteriori} assessment of $(a\sim c)$ mean temperature profile, $(d\sim f)$ wall-normal heat flux for VNC; here, inner scaling  \citep{george1979theory,ng2013turbulent} is used. $\solid$ GEP, {\color{gray} $\dotted$} DNS, {\color{gray} $\dashed$} baseline.}
		\label{fig:vn_prof}
	\end{center}
\end{figure*}
\begin{figure*}[!ht]
	\begin{center}
		\includegraphics[width=1\textwidth]{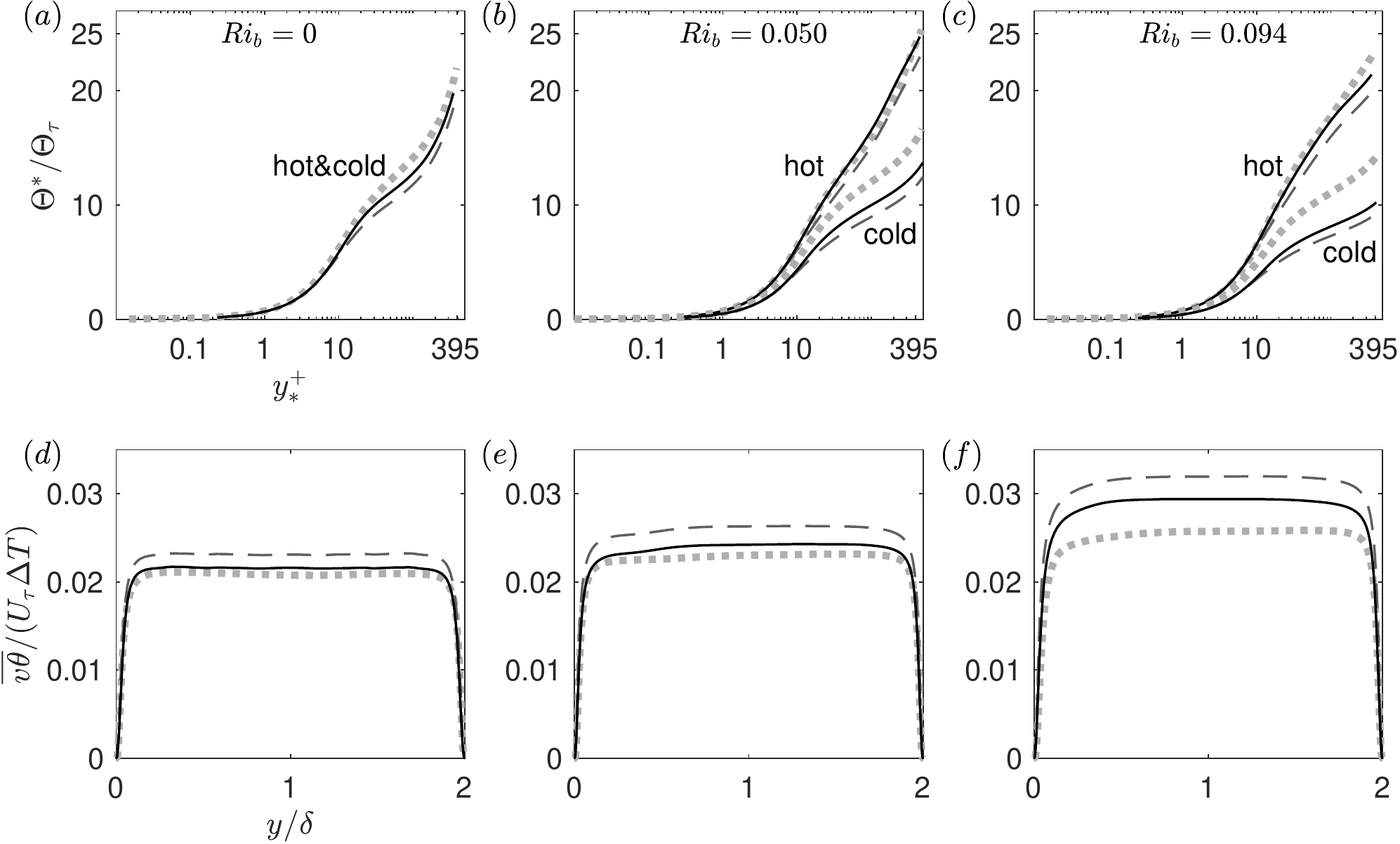}
		\caption{{A posteriori} assessment of $(a\sim c)$ mean temperature profile (at the hotter wall, $\Theta^* = \Theta_c-\Theta, y^+_* = y^+_h$;  at the colder wall, $\Theta^* = \Theta -\Theta_c, y^+_* = y^+_c$.  It is normalized by $\Theta_\tau = {{\mathrm{d} \Theta}/{\mathrm{d} y}|_{w}\:\alpha/U_\tau}$, where $\alpha= {1}/({Re _\tau Pr})$), and $(d\sim f)$ wall-normal heat flux profile for VMC. $\solid$ GEP, {\color{gray} $\dotted$} DNS, {\color{gray} $\dashed$} baseline. }
		\label{fig:mx_prof}
	\end{center}
\end{figure*}

Further results for the mean profile and wall-normal heat flux are shown in Fig.~\ref{fig:vn_prof} for VNC and Fig.~\ref{fig:mx_prof} for VMC. 
In Fig.~\ref{fig:vn_prof}, $(a\sim f)$ are plotted with an inner layer scaling \citep{george1979theory,ng2013turbulent}, where the inner temperature scales ${\left[|f_w|^3/(g\beta\kappa)\right]^{1/4}}$ and the inner length scale ${\left[\kappa^3/(g\beta |f_w|)\right]^{1/4}}$, in order to show the near-wall results. Due to the asymmetry profiles, here, we only show the half channel from the hotter wall. It is clear that the GEP model is better than the baseline for $\Theta$ and $\overline{v \theta}$ at the bulk region. 
In total, GEP-based models are fairly good, and the improvement holds for the whole field for VNC. 
Moreover, for VMC, 
Fig.~\ref{fig:mx_prof}$(a \sim c)$ shows the mean temperature $\Theta$ profile along the hotter and colder wall, respectively, and Fig.~\ref{fig:mx_prof}$(d \sim f)$ depicts the wall-normal heat flux $\overline{v \theta}$ in the global coordinates. 
Compared with the VNC case, the performance of baseline models is better, yet there is still room to improve. 
Although the baseline model correctly predicts the $Nu$ in the forced convection case, surprisingly, the improvement of GEP models on $\Theta$ and $\overline{v\theta}$  is consistently better than the baseline at different $Ri_b$ ($ 0\le Ri_b \le 0.094$).

\section{Concluding remarks}
\label{sec:con}
As the angle between the gravitation direction and temperature gradient reaches $90^\circ$, the turbulent Prandtl number $Pr_{t}$ and eddy viscosity $\nu_t$ tend to infinity in a thin adjustment region between the near-wall laminar-viscosity layer and the bulk turbulent region for vertical natural convection (VNC) in a range of Rayleigh numbers ($10^5\sim10^9$).
Whereas recent studies on VNC adopt an inner-outer two-layer structure \citep{george1979theory, holling2005asymptotic, ng2013turbulent, ng2015vertical}, we argue that this extra adjustment region can be identified by the zero point of Reynolds shear stress and the mean velocity gradient. 
Meanwhile, for vertical mixed convection (VMC) it also exists singular points of $Pr_{t}$ and $\nu_t$. 
They vary with increase of the buoyancy force, as the mean velocity maximum shift from the centreline  to the hotter wall.
This finding indicates that the primary effect of buoyancy on the mean profile for VMC is the break of symmetry, even for the flow in the shear-dominated regime ($0<Ri_b<0.1$).

To approximate the essential thermal quantities, including Nusselt number, mean temperature and wall-normal heat flux, we implement the machine learning framework via gene expression programming (GEP) to develop new turbulent heat flux models by using the DNS-based velocity fields as input for turbulent natural and mixed convection in a vertical channel.
Furthermore, a sensitivity study on the training dataset and cost functions via both \emph{frozen}  and \emph{CFD-driven} concepts are implemented to find the best prediction of the Nusselt number, mean temperature, and wall-normal heat flux.   
Comparing the \emph{a posteriori} performance on $Nu$, $\Theta$, and $\overline{v\theta}$, we discover that the error of the baseline (a constant $Pr_t = 0.90$) model for VNC case is larger than for the VMC case, and it is relatively easy to find effective GEP models for VNC. 
The data-driven method  in this study is almost independent of the training dataset and cost function for the VNC case. 
In contrast, the VMC cases needs a strict selection of both the training dataset and cost functions.
We discover that the inclusion of the mean gradient, which acts as a bridge between first- and second-order statistics, {in the cost function} shows significant advantages in finding a better GEP model. 
This is also true for the VNC cases.
In general, using cost functions that include the mean temperature gradient  based on the middle range of the DNS dataset across the parameter space for both VNC and  VMC can obtain a better model. 

The best performing GEP models can predict $Nu$ within a $5\%$ absolute percentage error for the VNC case across four decades of $Ra$ ($10^5\sim10^9$) and for VMC in the entire range of $0<Ri_b<0.1$ at a mean friction Reynolds number of 395, even though the training is carried out for a specific DNS dataset case. 
The reduction of error by GEP models is achieved across the current parameter space and cover all of the domain without any regional treatment.
It is also important that 
the data-driven method overcomes the singularity issues of linear gradient-based models with a spatially varying $Pr_t$.

The RANS model development is an odyssey when the pursuit is generality and universality. 
Nevertheless, we can show the robustness and accuracy of the current GEP models {for the turbulent Prandtl number}. 
We capture the correct physics of the turbulent Prandtl number, but concede that the result of full RANS-based CFD for VNC and VMC would still benefit from further improvement. 
One avenue to pursue in future work is addressing the fact that the time and length scale calculated by $k$ and $\varepsilon$ (or $\omega$) in RANS have a large discrepancy with the DNS dataset in the near-wall region, which undermines the suitability of  the  dimensionless velocity and temperature invariants.

\addcontentsline{toc}{section}{Declaration of Competing Interest}
\section*{Declaration of Competing Interest}
We wish to confirm that there are no known conflicts of interest associated with this publication.

\addcontentsline{toc}{section}{Acknowledgements}
\section*{Acknowledgements}
Xiaowei Xu was supported by the China Scholarship Council (No. 201606260057). We are indebted to Dr. Henry S. Dol, Dr. Chong Shen Ng, and Dr. Duncan Sutherland for providing the post-processed DNS data.

\addcontentsline{toc}{section}{Appendix A, DNS simulations on vertical mixed convection}
\section*{Appendix A, DNS  on vertical mixed convection} 
\label{sec:app}
The cited turbulent mixed convection cases are conducted by \cite{sutherland2015law}. As the reference is an abstract for the American Physical Society (APS) conference, it lacks computational details.  Therefore, after private communication, we can provide the simulation setup and governing equations. As previously mentioned, some flow parameters are held constant for all simulations. $Pr=0.709$, $\Delta \Theta=1$, and ${g_1}\beta=(-1,0,0)$. All simulations are carried out with computational domain size $(L_x, L_y, L_z) = (16h, 2h, 8h)$. The present grid spacing is uniform in the $x$- and $z$-directions and is stretched by a non-uniform Chebyshev grid $y_j=L_y \cos(\pi j/N_y)/2$ in the $y$-direction to resolve the steep, near-wall gradients.  The number of grid points $N_x$, $N_y$, and $N_z$ are chosen following \citet{kmm1987}, so that the $\Delta x^+ \approx 10$, $\Delta y^+ \approx 0.05$ and $\Delta z^+ \approx 5$ and to maintain an aspect ratio of approximately one in the centre of the channel. The time step is chosen to satisfy the CFL condition.
\begin{linenomath}
	\begin{align}
		&	\frac{\partial u_i}{\partial x_i}=0,\\ 
		&	\frac{\partial u_i}{\partial t} + u_j \frac{\partial u_i}{\partial x_j} =
		-\frac{1}{\rho_0}\frac{\partial p}{\partial x_i} + \delta_{i1} g_1 \beta \Theta + \nu \frac{\partial^2 u_i}{\partial x_j^2},\\		 
		&	\frac{\partial \Theta}{\partial t} + u_j\frac{\partial \Theta}{\partial x_j}=\kappa\frac{\partial^2 \Theta}{\partial x_j^2}.
	\end{align}
\end{linenomath}
\begin{table*}[!ht]
	\centering
	\begin{tabular*}{1.\textwidth}{@{\extracolsep{\fill}}llcccccccccc}
		\toprule
		label & Flow case   &    $Re_b$ & $Ra$   & $Ri_b$    &  $Re^g_{\tau}$        & $Re^h_{\tau}$ & $Re^c_{\tau}$ & $Nu$ & $N_x$ & $N_y$ & $N_z$ \\
		\hline
		Ri00 & Ra0\_Re4.1        & 13846    &   0     & 0        &  395.33            & 395.37 & 395.28  &  12.75    & 512  & 256   &  256  \\  
		Ri13 & Ra6.3\_Re4.2      & 14239    & $1.9\times10^6$  & $0.013$      &  405.33    & 418.86 & 391.80  &  12.79     & 512  & 256   &  256  \\  
		Ri18 & Ra6.3\_Re4.1      & 12963    & $2.2\times10^6$  & $0.018$      &  375.19     & 391.45  & 358.94  &  12.32     & 512  & 256   &  256  \\  
		Ri23 & Ra6.5\_Re4.2        &  14710    & $3.6\times10^6$   & $0.023$    &  419.81     & 440.46  & 399.16  &  13.57    & 512  & 384   &  256  \\  
		Ri35 & Ra6.6\_Re4.1        &  12696    & $4.0\times10^6$   & $0.035$    &  370.92      & 396.04  & 345.79  &  12.86      & 512  & 384   &  256  \\   
		Ri50 & Ra6.9\_Re4.2      &  15232    & $8.3\times10^6$  & $0.050$     &  438.10      & 484.09  & 392.11 &  14.88    & 512  & 384   &  256  \\  
		Ri94 & Ra7.0\_Re4.1      &  11825    & $9.3\times10^6$  & $0.094$     &  356.11     & 398.27  & 313.95 &  13.54    & 512  & 384   &  256  \\  
		\bottomrule
	\end{tabular*}
	\caption{List of parameters for a vertical buoyant turbulent channel. $Re_b = 2 h U_b / \nu$ is the bulk Reynolds number, $Re^g_{\tau} = h U_{\tau} / \nu$ is the friction Reynolds number, $Ri_b=2 \beta g \Delta \Theta h / U_b^2$ is the bulk Richardson number, $Ra=\beta g \Delta \Theta (2 h)^3 / (\alpha \nu)$ is the Rayleigh number, $Nu= (2h/\Delta\Theta)|{{\mathrm{d} \Theta}/{\mathrm{d} y}}|_w$ is the Nusselt number. $N_x$, $N_y$, $N_z$ are the number of grid points in the streamwise, wall-normal, and spanwise directions, respectively. The grid is stretched using a Chebyshev grid. }  
	\label{tab:params}
\end{table*}

The numerical scheme used is a fully conservative fourth-order finite difference method on a staggered grid for the velocities following \citet{morinishi1998}, and the temperature field is advected using the QUICK scheme \citep{leonard1979}. Time-stepping is accomplished by a low-storage third-order Runge-Kutta scheme due to \citet{spalart1991}. The continuity equation is enforced using the time-splitting method \citep{kimmoin1985}. The solver has been successfully used for some recent studies, for example, \cite{chung2012direct} and \citet{ng2015vertical}.

\bibliographystyle{unsrtnat}




\begin{thebibliography}{50}
	\expandafter\ifx\csname natexlab\endcsname\relax\def\natexlab#1{#1}\fi
	\providecommand{\url}[1]{\texttt{#1}}
	\providecommand{\href}[2]{#2}
	\providecommand{\path}[1]{#1}
	\providecommand{\DOIprefix}{doi:}
	\providecommand{\ArXivprefix}{arXiv:}
	\providecommand{\URLprefix}{URL: }
	\providecommand{\Pubmedprefix}{pmid:}
	\providecommand{\doi}[1]{\href{http://dx.doi.org/#1}{\path{#1}}}
	\providecommand{\Pubmed}[1]{\href{pmid:#1}{\path{#1}}}
	\providecommand{\bibinfo}[2]{#2}
	\ifx\xfnm\relax \def\xfnm[#1]{\unskip,\space#1}\fi
	\bibitem[{Batchelor(1954)}]{batchelor1954heat}
	\bibinfo{author}{Batchelor, G.}, \bibinfo{year}{1954}.
	\newblock \bibinfo{title}{Heat transfer by free convection across a closed
		cavity between vertical boundaries at different temperatures}.
	\newblock \bibinfo{journal}{Q. Appl. Math.} \bibinfo{volume}{12},
	\bibinfo{pages}{209--233}.
	\bibitem[{Boudjemadi et~al.(1997)Boudjemadi, Maupu, Laurence and
		Qu{\'e}r{\'e}}]{boudjemadi1997budgets}
	\bibinfo{author}{Boudjemadi, R.}, \bibinfo{author}{Maupu, V.},
	\bibinfo{author}{Laurence, D.}, \bibinfo{author}{Qu{\'e}r{\'e}, P.L.},
	\bibinfo{year}{1997}.
	\newblock \bibinfo{title}{Budgets of turbulent stresses and fluxes in a
		vertical slot natural convection flow at rayleigh {R}a$= 10^5$ and $5.4
		\times 10^5$}.
	\newblock \bibinfo{journal}{Int. J. Heat Fluid Flow} \bibinfo{volume}{18},
	\bibinfo{pages}{70--79}.
	\bibitem[{Chung and Matheou(2012)}]{chung2012direct}
	\bibinfo{author}{Chung, D.}, \bibinfo{author}{Matheou, G.},
	\bibinfo{year}{2012}.
	\newblock \bibinfo{title}{Direct numerical simulation of stationary homogeneous
		stratified sheared turbulence}.
	\newblock \bibinfo{journal}{J. Fluid Mech.} \bibinfo{volume}{696},
	\bibinfo{pages}{434--467}.
	\bibitem[{Dol et~al.(1999)Dol, Hanjali{\'c} and Versteegh}]{dol1999dns}
	\bibinfo{author}{Dol, H.}, \bibinfo{author}{Hanjali{\'c}, K.},
	\bibinfo{author}{Versteegh, T.}, \bibinfo{year}{1999}.
	\newblock \bibinfo{title}{A {DNS}-based thermal second-moment closure for
		buoyant convection at vertical walls}.
	\newblock \bibinfo{journal}{J. Fluid Mech.} \bibinfo{volume}{391},
	\bibinfo{pages}{211--247}.
	\bibitem[{Duraisamy et~al.(2019)Duraisamy, Iaccarino and
		Xiao}]{duraisamy2019turbulence}
	\bibinfo{author}{Duraisamy, K.}, \bibinfo{author}{Iaccarino, G.},
	\bibinfo{author}{Xiao, H.}, \bibinfo{year}{2019}.
	\newblock \bibinfo{title}{Turbulence modeling in the age of data}.
	\newblock \bibinfo{journal}{Annu. Rev. Fluid Mech.} \bibinfo{volume}{51},
	\bibinfo{pages}{357--377}.
	\bibitem[{Fabregat et~al.(2010)Fabregat, Pallares, Vernet, Cuesta, Ferr{\'e}
		and Grau}]{fabregat2010identification}
	\bibinfo{author}{Fabregat, A.}, \bibinfo{author}{Pallares, J.},
	\bibinfo{author}{Vernet, A.}, \bibinfo{author}{Cuesta, I.},
	\bibinfo{author}{Ferr{\'e}, J.}, \bibinfo{author}{Grau, F.},
	\bibinfo{year}{2010}.
	\newblock \bibinfo{title}{Identification of near-wall flow structures producing
		large wall transfer rates in turbulent mixed convection channel flow}.
	\newblock \bibinfo{journal}{Comput. Fluids} \bibinfo{volume}{39},
	\bibinfo{pages}{15--24}.
	\bibitem[{Garcia-Villalba and del Alamo(2011)}]{garcia2011turbulence}
	\bibinfo{author}{Garcia-Villalba, M.}, \bibinfo{author}{del Alamo, J.C.},
	\bibinfo{year}{2011}.
	\newblock \bibinfo{title}{Turbulence modification by stable stratification in
		channel flow}.
	\newblock \bibinfo{journal}{Phys. Fluids} \bibinfo{volume}{23},
	\bibinfo{pages}{045104}.
	\bibitem[{George~Jr and Capp(1979)}]{george1979theory}
	\bibinfo{author}{George~Jr, W.K.}, \bibinfo{author}{Capp, S.P.},
	\bibinfo{year}{1979}.
	\newblock \bibinfo{title}{A theory for natural convection turbulent boundary
		layers next to heated vertical surfaces}.
	\newblock \bibinfo{journal}{Int. J. Heat Mass Transfer} \bibinfo{volume}{22},
	\bibinfo{pages}{813--826}.
	\bibitem[{Gibson and Launder(1978)}]{gibson1978ground}
	\bibinfo{author}{Gibson, M.}, \bibinfo{author}{Launder, B.},
	\bibinfo{year}{1978}.
	\newblock \bibinfo{title}{Ground effects on pressure fluctuations in the
		atmospheric boundary layer}.
	\newblock \bibinfo{journal}{J. Fluid Mech.} \bibinfo{volume}{86},
	\bibinfo{pages}{491--511}.
	\bibitem[{Gibson and Leslie(1984)}]{gibson1984turbulent}
	\bibinfo{author}{Gibson, M.}, \bibinfo{author}{Leslie, D.},
	\bibinfo{year}{1984}.
	\newblock \bibinfo{title}{The turbulent {P}randtl number in the flow near a
		heated vertical surface}.
	\newblock \bibinfo{journal}{Int. Commun. Heat Mass Transf.}
	\bibinfo{volume}{11}, \bibinfo{pages}{73--84}.
	\bibitem[{Hanjali{\'c}(2002)}]{hanjalic2002one}
	\bibinfo{author}{Hanjali{\'c}, K.}, \bibinfo{year}{2002}.
	\newblock \bibinfo{title}{One-point closure models for buoyancy-driven
		turbulent flows}.
	\newblock \bibinfo{journal}{Annu. Rev. Fluid Mech.} \bibinfo{volume}{34},
	\bibinfo{pages}{321--347}.
	\bibitem[{H{\"o}lling and Herwig(2005)}]{holling2005asymptotic}
	\bibinfo{author}{H{\"o}lling, M.}, \bibinfo{author}{Herwig, H.},
	\bibinfo{year}{2005}.
	\newblock \bibinfo{title}{Asymptotic analysis of the near-wall region of
		turbulent natural convection flows}.
	\newblock \bibinfo{journal}{J. Fluid Mech.} \bibinfo{volume}{541},
	\bibinfo{pages}{383}.
	\bibitem[{Jackson et~al.(1989)Jackson, Cotton and Axcell}]{jackson1989studies}
	\bibinfo{author}{Jackson, J.}, \bibinfo{author}{Cotton, M.},
	\bibinfo{author}{Axcell, B.}, \bibinfo{year}{1989}.
	\newblock \bibinfo{title}{Studies of mixed convection in vertical tubes}.
	\newblock \bibinfo{journal}{Int. J. Heat Fluid Flow} \bibinfo{volume}{10},
	\bibinfo{pages}{2--15}.
	\bibitem[{Kasagi and Nishimura(1997)}]{kasagi1997direct}
	\bibinfo{author}{Kasagi, N.}, \bibinfo{author}{Nishimura, M.},
	\bibinfo{year}{1997}.
	\newblock \bibinfo{title}{Direct numerical simulation of combined forced and
		natural turbulent convection in a vertical plane channel}.
	\newblock \bibinfo{journal}{Int. J. Heat Fluid Flow} \bibinfo{volume}{18},
	\bibinfo{pages}{88--99}.
	\bibitem[{Kays(1994)}]{kays1994turbulent}
	\bibinfo{author}{Kays, W.M.}, \bibinfo{year}{1994}.
	\newblock \bibinfo{title}{Turbulent {P}randtl number, where are we?}
	\newblock \bibinfo{journal}{J. Heat Transfer} \bibinfo{volume}{116},
	\bibinfo{pages}{284--295}.
	\bibitem[{Kim and Moin(1985)}]{kimmoin1985}
	\bibinfo{author}{Kim, J.}, \bibinfo{author}{Moin, P.}, \bibinfo{year}{1985}.
	\newblock \bibinfo{title}{Application of a fractional-step method to
		incompressible {N}avier-{S}tokes equations}.
	\newblock \bibinfo{journal}{J. Comput. Phys.} \bibinfo{volume}{59},
	\bibinfo{pages}{308--323}.
	\bibitem[{Kim et~al.(1987)Kim, Moin and Moser}]{kmm1987}
	\bibinfo{author}{Kim, J.}, \bibinfo{author}{Moin, P.}, \bibinfo{author}{Moser,
		R.}, \bibinfo{year}{1987}.
	\newblock \bibinfo{title}{Turbulence statistics in fully developed channel flow
		at low {R}eynolds number}.
	\newblock \bibinfo{journal}{J. Fluid Mech.} \bibinfo{volume}{177},
	\bibinfo{pages}{133--166}.
	\bibitem[{Ki{\v{s}} and Herwig(2014)}]{kivs2014natural}
	\bibinfo{author}{Ki{\v{s}}, P.}, \bibinfo{author}{Herwig, H.},
	\bibinfo{year}{2014}.
	\newblock \bibinfo{title}{Natural convection in a vertical plane channel: {DNS}
		results for high {G}rashof numbers}.
	\newblock \bibinfo{journal}{Heat and Mass Transfer} \bibinfo{volume}{50},
	\bibinfo{pages}{957--972}.
	\bibitem[{Kutz(2017)}]{kutz2017deep}
	\bibinfo{author}{Kutz, J.N.}, \bibinfo{year}{2017}.
	\newblock \bibinfo{title}{Deep learning in fluid dynamics}.
	\newblock \bibinfo{journal}{J. Fluid Mech.} \bibinfo{volume}{814},
	\bibinfo{pages}{1--4}.
	\bibitem[{Leonard(1979)}]{leonard1979}
	\bibinfo{author}{Leonard, B.P.}, \bibinfo{year}{1979}.
	\newblock \bibinfo{title}{A stable and accurate convective modelling procedure
		based on quadratic upstream interpolation}.
	\newblock \bibinfo{journal}{Comput. Methods Appl. Mech. Eng.}
	\bibinfo{volume}{19}, \bibinfo{pages}{59--98}.
	\bibitem[{Li(2019)}]{li2019turbulent}
	\bibinfo{author}{Li, D.}, \bibinfo{year}{2019}.
	\newblock \bibinfo{title}{Turbulent {P}randtl number in the atmospheric
		boundary layer-where are we now?}
	\newblock \bibinfo{journal}{Atmos. Res.} \bibinfo{volume}{216},
	\bibinfo{pages}{86--105}.
	\bibitem[{Li et~al.(2015)Li, Katul and Zilitinkevich}]{li2015revisiting}
	\bibinfo{author}{Li, D.}, \bibinfo{author}{Katul, G.G.},
	\bibinfo{author}{Zilitinkevich, S.S.}, \bibinfo{year}{2015}.
	\newblock \bibinfo{title}{Revisiting the turbulent {P}randtl number in an
		idealized atmospheric surface layer}.
	\newblock \bibinfo{journal}{J. Atmos. Sci.} \bibinfo{volume}{72},
	\bibinfo{pages}{2394--2410}.
	\bibitem[{Ling et~al.(2016)Ling, Ryan, Bodart and Eaton}]{ling2016analysis}
	\bibinfo{author}{Ling, J.}, \bibinfo{author}{Ryan, K.J.},
	\bibinfo{author}{Bodart, J.}, \bibinfo{author}{Eaton, J.K.},
	\bibinfo{year}{2016}.
	\newblock \bibinfo{title}{Analysis of turbulent scalar flux models for a
		discrete hole film cooling flow}.
	\newblock \bibinfo{journal}{J. Turbomach.} \bibinfo{volume}{138},
	\bibinfo{pages}{011006}.
	\bibitem[{Mellor and Yamada(1982)}]{mellor1982development}
	\bibinfo{author}{Mellor, G.L.}, \bibinfo{author}{Yamada, T.},
	\bibinfo{year}{1982}.
	\newblock \bibinfo{title}{Development of a turbulence closure model for
		geophysical fluid problems}.
	\newblock \bibinfo{journal}{Rev. Geophys.} \bibinfo{volume}{20},
	\bibinfo{pages}{851--875}.
	\bibitem[{Milani et~al.(2020)Milani, Ling and Eaton}]{milani2020turbulent}
	\bibinfo{author}{Milani, P.M.}, \bibinfo{author}{Ling, J.},
	\bibinfo{author}{Eaton, J.K.}, \bibinfo{year}{2020}.
	\newblock \bibinfo{title}{Turbulent scalar flux in inclined jets in crossflow:
		counter gradient transport and deep learning modelling}.
	\newblock \bibinfo{journal}{arXiv preprint arXiv:2001.04600} .
	\bibitem[{Milani et~al.(2018)Milani, Ling, Saez-Mischlich, Bodart and
		Eaton}]{milani2018machine}
	\bibinfo{author}{Milani, P.M.}, \bibinfo{author}{Ling, J.},
	\bibinfo{author}{Saez-Mischlich, G.}, \bibinfo{author}{Bodart, J.},
	\bibinfo{author}{Eaton, J.K.}, \bibinfo{year}{2018}.
	\newblock \bibinfo{title}{A machine learning approach for determining the
		turbulent diffusivity in film cooling flows}.
	\newblock \bibinfo{journal}{J. Turbomach.} \bibinfo{volume}{140},
	\bibinfo{pages}{021006}.
	\bibitem[{Monin and Obukhov(1954)}]{monin1954basic}
	\bibinfo{author}{Monin, A.S.}, \bibinfo{author}{Obukhov, A.M.},
	\bibinfo{year}{1954}.
	\newblock \bibinfo{title}{Basic laws of turbulent mixing in the surface layer
		of the atmosphere}.
	\newblock \bibinfo{journal}{Contrib. Geophys. Inst. Acad. Sci. USSR}
	\bibinfo{volume}{151}, \bibinfo{pages}{e187}.
	\bibitem[{Morinishi et~al.(1998)Morinishi, Lund, Vasilyev and
		Moin}]{morinishi1998}
	\bibinfo{author}{Morinishi, Y.}, \bibinfo{author}{Lund, T.},
	\bibinfo{author}{Vasilyev, O.}, \bibinfo{author}{Moin, P.},
	\bibinfo{year}{1998}.
	\newblock \bibinfo{title}{Fully conservative higher order finite difference
		schemes for incompressible flow}.
	\newblock \bibinfo{journal}{J. Comput. Phys.} \bibinfo{volume}{143},
	\bibinfo{pages}{90--124}.
	\bibitem[{Myong and Kasagi(1990)}]{myong1990new}
	\bibinfo{author}{Myong, H.K.}, \bibinfo{author}{Kasagi, N.},
	\bibinfo{year}{1990}.
	\newblock \bibinfo{title}{A new approach to the improvement of $k-\varepsilon$
		turbulence model for wall-bounded shear flows}.
	\newblock \bibinfo{journal}{JSME international journal. Ser. 2, Fluids
		engineering, heat transfer, power, combustion, thermophysical properties}
	\bibinfo{volume}{33}, \bibinfo{pages}{63--72}.
	\bibitem[{Ng et~al.(2013)Ng, Chung and Ooi}]{ng2013turbulent}
	\bibinfo{author}{Ng, C.}, \bibinfo{author}{Chung, D.}, \bibinfo{author}{Ooi,
		A.}, \bibinfo{year}{2013}.
	\newblock \bibinfo{title}{Turbulent natural convection scaling in a vertical
		channel}.
	\newblock \bibinfo{journal}{Int. J. Heat Fluid Flow} \bibinfo{volume}{44},
	\bibinfo{pages}{554--562}.
	\bibitem[{Ng et~al.(2015)Ng, Ooi, Lohse and Chung}]{ng2015vertical}
	\bibinfo{author}{Ng, C.S.}, \bibinfo{author}{Ooi, A.}, \bibinfo{author}{Lohse,
		D.}, \bibinfo{author}{Chung, D.}, \bibinfo{year}{2015}.
	\newblock \bibinfo{title}{Vertical natural convection: application of the
		unifying theory of thermal convection}.
	\newblock \bibinfo{journal}{J. Fluid Mech.} \bibinfo{volume}{764},
	\bibinfo{pages}{349--361}.
	\bibitem[{Ng et~al.(2017)Ng, Ooi, Lohse and Chung}]{ng2017changes}
	\bibinfo{author}{Ng, C.S.}, \bibinfo{author}{Ooi, A.}, \bibinfo{author}{Lohse,
		D.}, \bibinfo{author}{Chung, D.}, \bibinfo{year}{2017}.
	\newblock \bibinfo{title}{Changes in the boundary-layer structure at the edge
		of the ultimate regime in vertical natural convection}.
	\newblock \bibinfo{journal}{J. Fluid Mech.} \bibinfo{volume}{825},
	\bibinfo{pages}{550--572}.
	\bibitem[{Phillips(1996)}]{phillips1996direct}
	\bibinfo{author}{Phillips, J.}, \bibinfo{year}{1996}.
	\newblock \bibinfo{title}{Direct simulations of turbulent unstratified natural
		convection in a vertical slot for {Pr}= 0.71}.
	\newblock \bibinfo{journal}{Int. J. Heat Mass Transfer} \bibinfo{volume}{39},
	\bibinfo{pages}{2485--2494}.
	\bibitem[{Pirozzoli et~al.(2017)Pirozzoli, Bernardini, Verzicco and
		Orlandi}]{pirozzoli2017mixed}
	\bibinfo{author}{Pirozzoli, S.}, \bibinfo{author}{Bernardini, M.},
	\bibinfo{author}{Verzicco, R.}, \bibinfo{author}{Orlandi, P.},
	\bibinfo{year}{2017}.
	\newblock \bibinfo{title}{Mixed convection in turbulent channels with unstable
		stratification}.
	\newblock \bibinfo{journal}{J. Fluid Mech.} \bibinfo{volume}{821},
	\bibinfo{pages}{482--516}.
	\bibitem[{Reynolds(1975)}]{reynolds1975prediction}
	\bibinfo{author}{Reynolds, A.}, \bibinfo{year}{1975}.
	\newblock \bibinfo{title}{The prediction of turbulent {P}randtl and {S}chmidt
		numbers}.
	\newblock \bibinfo{journal}{Int. J. Heat Mass Transfer} \bibinfo{volume}{18},
	\bibinfo{pages}{1055--1069}.
	\bibitem[{Rodi and Mansour(1993)}]{rodi1993low}
	\bibinfo{author}{Rodi, W.}, \bibinfo{author}{Mansour, N.},
	\bibinfo{year}{1993}.
	\newblock \bibinfo{title}{Low reynolds number $k-\varepsilon$ modelling with
		the aid of direct simulation data}.
	\newblock \bibinfo{journal}{J. Fluid Mech.} \bibinfo{volume}{250},
	\bibinfo{pages}{509--529}.
	\bibitem[{Sandberg et~al.(2018)Sandberg, Tan, Weatheritt, Ooi, Haghiri,
		Michelassi and Laskowski}]{sandberg2018applying}
	\bibinfo{author}{Sandberg, R.}, \bibinfo{author}{Tan, R.},
	\bibinfo{author}{Weatheritt, J.}, \bibinfo{author}{Ooi, A.},
	\bibinfo{author}{Haghiri, A.}, \bibinfo{author}{Michelassi, V.},
	\bibinfo{author}{Laskowski, G.}, \bibinfo{year}{2018}.
	\newblock \bibinfo{title}{Applying machine learnt explicit algebraic stress and
		scalar flux models to a fundamental trailing edge slot}.
	\newblock \bibinfo{journal}{J. Turbomach.} \bibinfo{volume}{140},
	\bibinfo{pages}{101008}.
	\bibitem[{Shih and Lumley(1993)}]{shih1993remarks}
	\bibinfo{author}{Shih, T.H.}, \bibinfo{author}{Lumley, J.L.},
	\bibinfo{year}{1993}.
	\newblock \bibinfo{title}{Remarks on turbulent constitutive relations}.
	\newblock \bibinfo{journal}{Math. comput. model.} \bibinfo{volume}{18},
	\bibinfo{pages}{9--16}.
	\bibitem[{Spalart et~al.(1991)Spalart, Moser and Rogers}]{spalart1991}
	\bibinfo{author}{Spalart, P.R.}, \bibinfo{author}{Moser, R.D.},
	\bibinfo{author}{Rogers, M.M.}, \bibinfo{year}{1991}.
	\newblock \bibinfo{title}{Spectral methods for the {N}avier-{S}tokes equations
		with one infinite and two periodic directions}.
	\newblock \bibinfo{journal}{J. Comput. Phys.} \bibinfo{volume}{96},
	\bibinfo{pages}{297--324}.
	\bibitem[{Sutherland et~al.(2015)Sutherland, Chung, Ooi and
		Bou-Zeid}]{sutherland2015law}
	\bibinfo{author}{Sutherland, D.}, \bibinfo{author}{Chung, D.},
	\bibinfo{author}{Ooi, A.}, \bibinfo{author}{Bou-Zeid, E.},
	\bibinfo{year}{2015}.
	\newblock \bibinfo{title}{The law-of-the-wall in mixed convection flow in a
		vertical channel}.
	\newblock \bibinfo{journal}{APS} , \bibinfo{pages}{A19--009}.
	\bibitem[{Versteegh and Nieuwstadt(1999)}]{versteegh1999direct}
	\bibinfo{author}{Versteegh, T.}, \bibinfo{author}{Nieuwstadt, F.},
	\bibinfo{year}{1999}.
	\newblock \bibinfo{title}{A direct numerical simulation of natural convection
		between two infinite vertical differentially heated walls scaling laws and
		wall functions}.
	\newblock \bibinfo{journal}{Int. J. Heat Mass Transfer} \bibinfo{volume}{42},
	\bibinfo{pages}{3673--3693}.
	\bibitem[{Weatheritt and Sandberg(2016)}]{weatheritt2016novel}
	\bibinfo{author}{Weatheritt, J.}, \bibinfo{author}{Sandberg, R.},
	\bibinfo{year}{2016}.
	\newblock \bibinfo{title}{A novel evolutionary algorithm applied to algebraic
		modifications of the {RANS} stress-strain relationship}.
	\newblock \bibinfo{journal}{J. Comput. Phys.} \bibinfo{volume}{325},
	\bibinfo{pages}{22--37}.
	\bibitem[{Weatheritt and Sandberg(2017)}]{weatheritt2017development}
	\bibinfo{author}{Weatheritt, J.}, \bibinfo{author}{Sandberg, R.},
	\bibinfo{year}{2017}.
	\newblock \bibinfo{title}{The development of algebraic stress models using a
		novel evolutionary algorithm}.
	\newblock \bibinfo{journal}{Int. J. Heat Fluid Flow} \bibinfo{volume}{68},
	\bibinfo{pages}{298--318}.
	\bibitem[{Weatheritt et~al.(2017)Weatheritt, Sandberg, Ling, Saez and
		Bodart}]{weatheritt2017comparative}
	\bibinfo{author}{Weatheritt, J.}, \bibinfo{author}{Sandberg, R.D.},
	\bibinfo{author}{Ling, J.}, \bibinfo{author}{Saez, G.},
	\bibinfo{author}{Bodart, J.}, \bibinfo{year}{2017}.
	\newblock \bibinfo{title}{A comparative study of contrasting machine learning
		frameworks applied to {RANS} modeling of jets in crossflow}, in:
	\bibinfo{booktitle}{ASME Turbo Expo 2017: Turbomachinery Technical Conference
		and Exposition}, \bibinfo{organization}{American Society of Mechanical
		Engineers}. pp. \bibinfo{pages}{V02BT41A012--V02BT41A012}.
	\bibitem[{Weatheritt et~al.(2020)Weatheritt, Zhao, Sandberg, Mizukami and
		Tanimoto}]{weatheritt2020data}
	\bibinfo{author}{Weatheritt, J.}, \bibinfo{author}{Zhao, Y.},
	\bibinfo{author}{Sandberg, R.D.}, \bibinfo{author}{Mizukami, S.},
	\bibinfo{author}{Tanimoto, K.}, \bibinfo{year}{2020}.
	\newblock \bibinfo{title}{Data-driven scalar-flux model development with
		application to jet in cross flow}.
	\newblock \bibinfo{journal}{Int. J. Heat Fluid Flow} \bibinfo{volume}{147},
	\bibinfo{pages}{118931}.
	\bibitem[{Wells and Worster(2008)}]{wells2008geophysical}
	\bibinfo{author}{Wells, A.J.}, \bibinfo{author}{Worster, M.G.},
	\bibinfo{year}{2008}.
	\newblock \bibinfo{title}{A geophysical-scale model of vertical natural
		convection boundary layers}.
	\newblock \bibinfo{journal}{J. Fluid Mech.} \bibinfo{volume}{609},
	\bibinfo{pages}{111}.
	\bibitem[{Wetzel and Wagner(2019)}]{wetzel2019buoyancy}
	\bibinfo{author}{Wetzel, T.}, \bibinfo{author}{Wagner, C.},
	\bibinfo{year}{2019}.
	\newblock \bibinfo{title}{Buoyancy-induced effects on large-scale motions in
		differentially heated vertical channel flows studied in direct numerical
		simulations}.
	\newblock \bibinfo{journal}{Int. J. Heat Fluid Flow} \bibinfo{volume}{75},
	\bibinfo{pages}{14--26}.
	\bibitem[{Xu et~al.(1998)Xu, Chen and Nieuwstadt}]{xu1998new}
	\bibinfo{author}{Xu, W.}, \bibinfo{author}{Chen, Q.},
	\bibinfo{author}{Nieuwstadt, F.}, \bibinfo{year}{1998}.
	\newblock \bibinfo{title}{A new turbulence model for near-wall natural
		convection}.
	\newblock \bibinfo{journal}{Int. J. Heat Mass Transfer} \bibinfo{volume}{41},
	\bibinfo{pages}{3161--3176}.
	\bibitem[{Zhao et~al.(2020)Zhao, Akolekar, Weatheritt, Michelassi and
		Sandberg}]{zhao2020rans}
	\bibinfo{author}{Zhao, Y.}, \bibinfo{author}{Akolekar, H.D.},
	\bibinfo{author}{Weatheritt, J.}, \bibinfo{author}{Michelassi, V.},
	\bibinfo{author}{Sandberg, R.D.}, \bibinfo{year}{2020}.
	\newblock \bibinfo{title}{{RANS} turbulence model development using
		{CFD}-driven machine learning}.
	\newblock \bibinfo{journal}{J. Comput. Phys.} \bibinfo{volume}{411},
	\bibinfo{pages}{109413}.
	\bibitem[{Zilitinkevich et~al.(2013)Zilitinkevich, Elperin, Kleeorin,
		Rogachevskii and Esau}]{zilitinkevich2013hierarchy}
	\bibinfo{author}{Zilitinkevich, S.}, \bibinfo{author}{Elperin, T.},
	\bibinfo{author}{Kleeorin, N.}, \bibinfo{author}{Rogachevskii, I.},
	\bibinfo{author}{Esau, I.}, \bibinfo{year}{2013}.
	\newblock \bibinfo{title}{A hierarchy of energy-and flux-budget ({EFB})
		turbulence closure models for stably-stratified geophysical flows}.
	\newblock \bibinfo{journal}{Bound.-Layer Meteorol.} \bibinfo{volume}{146},
	\bibinfo{pages}{341--373}.
	
\end{thebibliography}

\end{document}